\documentclass[aps,prb,reprint,superscriptaddress,showpacs]{revtex4-1}
\usepackage{graphicx}                 
\usepackage{amsmath}
\usepackage{textcomp}
\usepackage{multirow} 
\begin{document}
\title{Ultrafast laser-triggered emission from hafnium carbide tips}
\author{Catherine Kealhofer}
\author{Seth M. Foreman}
\affiliation{Physics Department, Stanford University, Stanford, California 94305, USA}
\author{Stefan Gerlich}
\affiliation{Vienna Center for Quantum Science and Technology (VCQ), University of Vienna, Faculty of Physics, Boltzmanngasse 5, 1090 Vienna, Austria}
\affiliation{Physics Department, Stanford University, Stanford, California 94305, USA}
\author{Mark A. Kasevich}
\affiliation{Physics Department, Stanford University, Stanford, California 94305, USA}
\date{\today}
\begin{abstract}
Electron emission from hafnium carbide (HfC) field emission tips induced by a sub-10~fs, 150~MHz repetition rate Ti:sapphire laser is studied.  Two-photon emission is observed at low power with a moderate electric bias field applied to the tips.  As the bias field and/or laser power is increased, the average current becomes dominated by thermally-enhanced field emission due to laser heating: both the low thermal conductivity of HfC and the laser's high repetition rate can lead to a temperature rise of several hundred Kelvin at the tip apex.  The contribution of current from a thermal transient at times shorter than the electron-phonon coupling time is considered in the context of the two-temperature model.  Under the conditions of this experiment, the integrated current from the thermal transient is shown to be negligible in comparison with the two-photon emission.  A finite element model of the laser heating and thermal conduction supports these conclusions and is also used to compare the nature of thermal effects in HfC, tungsten, and gold tips.
\end{abstract}

\maketitle
\section{Introduction}
Field emission tips are extremely high brightness emitters \cite{Spence94} with large transverse coherence length \cite{Cho04} and small energy spread \cite{Gomer61}.  They are therefore excellent sources for interferometry experiments such as electron holography \cite{Midgley09} and coherent electron diffraction. \cite{Zuo03}  Femtosecond field emission sources are being developed by triggering electron emission with ultrafast laser pulses. \cite{Hommelhoff06a, Hommelhoff06, Ropers07, Barwick07, Ganter08, Yanagisawa10, Yang10}  These sources have extremely small emission areas in comparison with existing ultrafast electron sources\cite{Zewail10, Miller10} based on flat photocathodes and should therefore lead to better resolution for diffraction and imaging. Furthermore, the small source size leads to the high transverse coherence required for electron interferometry and holography. \cite{Hasselbach10}   Ultrafast laser-triggered tips promise to extend the full range of electron microscopy imaging techniques to the femtosecond time-scale, creating tools for the study of chemical and physical transformations of materials with atomic-scale resolution.

A number of areas have already benefited from laser-triggered emission from sharp tips.  Optical field enhancement at the tip apex has allowed exploration of strong field physics in solid state systems without the need for laser amplifiers. \cite{Bormann10, Schenk10}  Due to the extreme sensitivity of current to the light intensity at the emission site, field emission tips have been used to study the temporal dynamics of the optical near-field induced by the tip itself \cite{Yanagisawa10} and map optical fields in nearby nano-structures \cite{Ropers07} with nanometric spatial resolution.  Ultrafast laser-triggered emission has also been used as a source of electron pulses for experiments that verified the dispersion-less nature of the Aharonov-Bohm effect \cite{Caprez07} and is being developed as an electron source for ultrafast scanning electron microscopy. \cite{Yang10}  These sources could also enable laser-based electron accelerators \cite{Plettner06} and a table-top ultrafast x-ray microfocus source.

Hafnium carbide (HfC) is an attractive material to use for ultrafast laser-induced electron emission because of its low work function ($\sim 3.4$~eV), which makes it easier to liberate electrons,  and its good mechanical stability at high temperatures. \cite{Kagarice08}  The low work function relaxes laser intensity requirements for electron emission, which would allow devices that use low power fiber lasers to excite the electrons or devices where electron optics elements near the tip preclude high numerical aperture focusing of the laser.  Here, we confirm this by showing that to generate one electron per pulse (an important regime for avoiding space charge while maintaining the maximum possible flux) HfC requires one quarter of the pulse energy as results reported in tungsten \cite{Schenk10} and gold. \cite{Bormann10}  HfC has been demonstrated to have higher emission current densities than tungsten in stable DC operation, \cite{Mackie93} and it may be annealed more aggressively than tungsten to create clean surfaces.  The low thermal conductivity of HfC is, however, a disadvantage if higher repetition frequencies or greater numbers of electrons per pulse are desired; it results in higher average temperatures and longer-lasting elevated temperature transients for pulsed laser heating in comparison with materials like gold or tungsten.

Several ultrafast laser-induced electron emission processes have been observed in field emission tips, and these in general transfer the precise timing of the laser pulses to the electrons.  In the perturbative regime, photo-assisted field emission \cite{Hommelhoff06a} and multi-photon emission \cite{Ropers07} are prompt with respect to the laser intensity envelope.  Work in tungsten \cite{Hommelhoff06, Schenk10} and gold \cite{Bormann10} has explored higher pulse fluences, where for sufficiently short optical pulses the emission profile is predicted to have sub-optical cycle features. \cite{Hommelhoff06, Schenk10}

Here, we show that laser-induced electron emission from HfC is due to two-photon emission over a broad range of laser power and bias field, yielding up to several electrons per pulse at 150~MHz. However, for certain parameters we observe thermally-enhanced field emission (TFE), which we describe as a transient current pulse (transient TFE) superimposed on a steady-state background (DC TFE).  DC TFE arises from a significantly elevated average apex temperature due to the low thermal conductivity of HfC and the laser's high repetition rate.  A transient TFE pulse could arise from rapid heating of the electrons by each laser pulse, followed by electron-phonon coupling on a ps timescale and cooling by conduction.  The sub-ns timescale of conduction following laser-induced heating is known to play a role in ultrafast laser-assisted field evaporation, \cite{Vurpillot09, Houard11} which is used for atom-probe tomography.  However, until now there have been no investigations of the effects of a laser-induced elevated electron temperature on the timing properties of electron emission from fs-laser-triggered tips.  By comparing the scaling of DC and transient TFE with that of multiphoton emission, we show that over the parameter range of our experiment, the transient effect is not a dominant source of current.

The paper is organized as follows.  The experimental setup is described in Section \ref{section:expsetup}, and Section \ref{section:emission_process} is an overview of emission processes.  Section \ref{section:two_photon_and_TFE} presents evidence for two-photon emission and DC thermal-assisted field emission as the dominant mechanisms observed in HfC.  Section \ref{section:modeling} describes modeling of laser heating of the tip.  In section \ref{section:transient_bounds}, these results are combined to place an upper bound on the fraction of pulsed current that is due to transient TFE and to compare how multiphoton emission and transient thermal-assisted field emission scale with pulse energy in a low repetition rate system.
\section{Experimental setup\label{section:expsetup}}
\begin{figure}
\includegraphics[width=8.6cm]{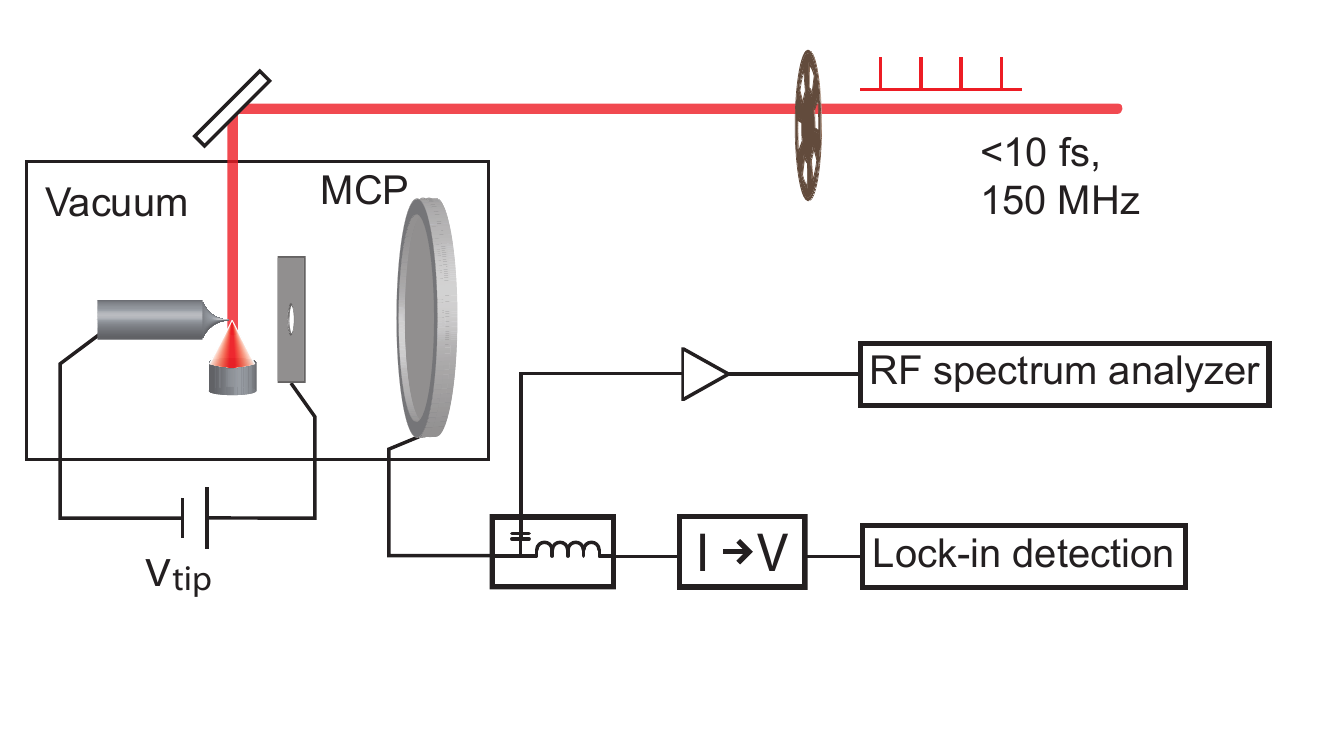}%
\caption{Experimental setup.  The ultrafast laser pulses are focused onto a HfC tip, which is maintained in ultra-high vacuum ($<5\times10^{-10}$~torr).  A bias voltage, $V_{\mathrm{tip}}$ is applied between the tip and extraction aperture, located 1.5 mm away.  On the other side of the aperture, the electrons propagate in a field free region.  Electrons are detected with an MCP; a bias tee splits the RF and near DC components of the current at the front of the MCP.  A chopper in the laser beam allows lock-in detection of $I_{laser}$.
 \label{expt}}
\end{figure}
A diagram of the experimental setup is shown in Fig. \ref{expt}.  The output of a sub-10~fs, 150~MHz repetition frequency Ti:sapphire oscillator is focused with a 2.7~\textmu m waist ($1/e^2$ intensity radius) onto a 310-oriented hafnium carbide field emission tip (Applied Physics Technologies, McMinnville, OR).  The dispersion of the vacuum window and other glass in the beam-line is compensated with double-chirped mirrors, and the $\sim10$~fs pulse duration at the tip is verified by an interferometric autocorrelation with electron emission from the tip as the non-linear detector. \cite{Hommelhoff06}  The tip has a 120~nm radius of curvature at the apex.  We take the work function of the tip's emitting surface to be 3.4 eV. \cite{Zaima84}  The tip is maintained in ultra-high vacuum ($<5\times10^{-10}$~torr).  Surface contamination modifies the work function of the tip; we therefore flash anneal to 2100~K for 1~s prior to each set of measurements.  A DC bias is applied between the tip and an extraction aperture, and the spatial distribution of the electrons is imaged using a dual chevron configuration microchannel plate (MCP) with a phosphor screen.  The absolute gain of the MCP was determined by counting detection events in high gain operation and then using a relative calibration for lower voltages across the plates.

A bias tee splits the RF and near-DC components of the current at the front (incident surface) of the MCP.  The DC component is used to measure the current induced by the laser ($I_{\mathrm{laser}}$) with lock-in detection at the frequency of a chopper in the optical beam path.  $I_{\mathrm{laser}}$ is equivalent to the difference between the average current incident upon the MCP with and without light on the tip.  We use $I_{\mathrm{DC}}$ to describe current from the tip when there is no laser illumination.  The RF signal contains harmonics of $f_{\mathrm{rep}}$, the laser repetition frequency, extending to frequencies $>2.5$~GHz, limited by the MCP's response.  Since the electron detection events have a known pulse shape, the amplitude of the harmonic at $2f_{\mathrm{rep}}$ can be converted into an equivalent average laser-induced current which we define as $I_{\mathrm{2f}}$.  For instance, when all the signal is pulsed (no appreciable DC TFE), $I_{\mathrm{2f}}$ and $I_{\mathrm{laser}}$ have the same value.

\section{Emission processes\label{section:emission_process}}
\begin{figure}
\includegraphics[width=8.6cm]{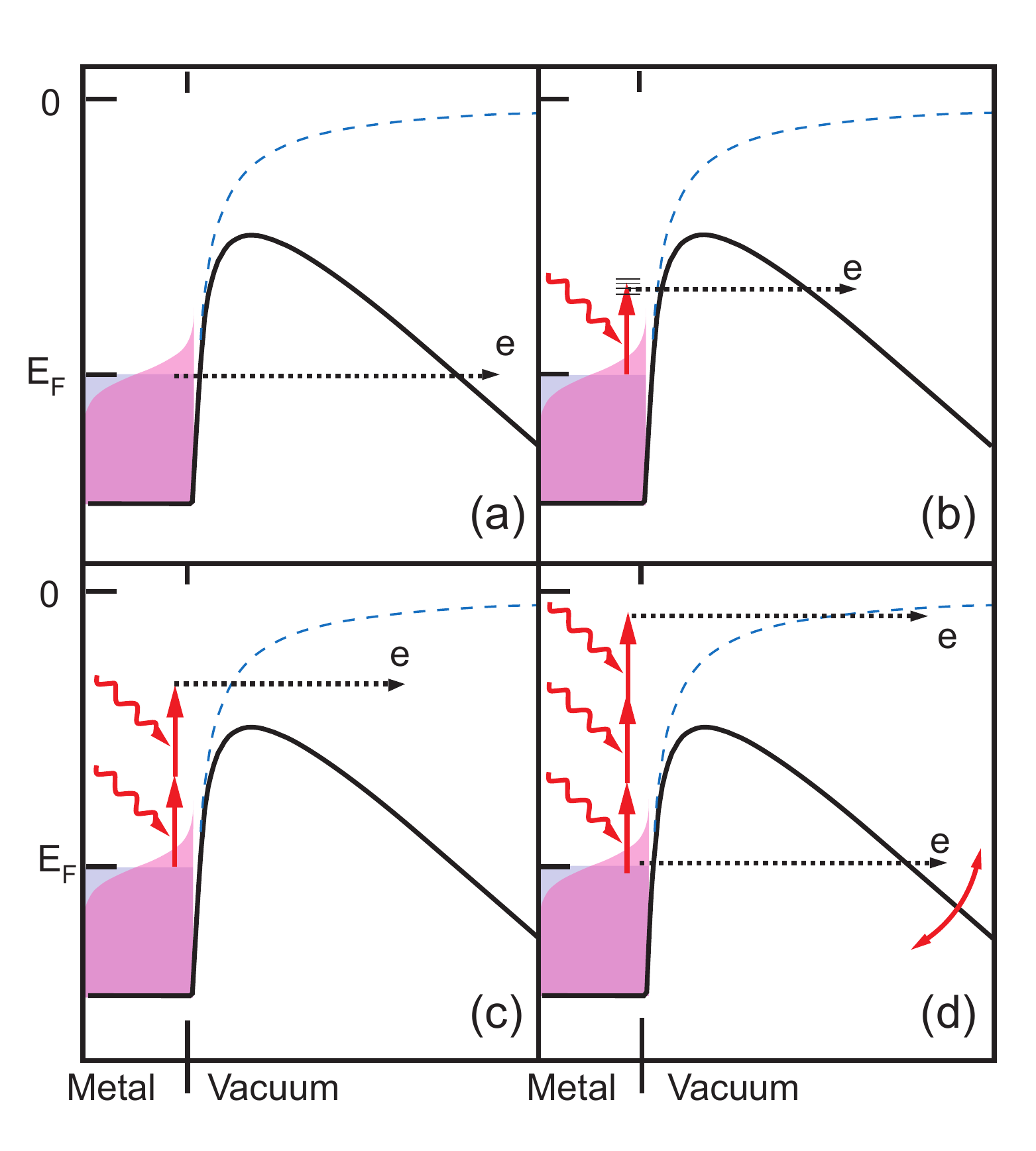}%
\caption{Electron emission processes.  (a) Field emission (including thermally-enhanced).  (b) Photo-assisted field emission.  (c) Multi-photon (over the barrier) emission.  (d) Above-threshold photoemission and optical field emission.  \label{emission_processes}}
\end{figure}

A diagram of DC and laser-induced electron emission processes is shown in Fig. \ref{emission_processes}.  We summarize each of these below.
\subsection{Field Emission\label{section:fieldemission}}
The field at the apex of a metal tip is given by $F = V/(kr)$, where $r$ is the radius of curvature at the apex of the tip and $k$ is a geometrical factor (typically $\sim$4-5).  For tips with $r$ of 10-100~nm, field strengths of GV/m may be obtained using modest voltages.  Under such large bias fields, electrons tunnel directly from the fermi level into the vacuum, as shown in Fig. \ref{emission_processes}(a).  Without illumination, the magnitude of the total current density emitted from a field emission tip as a function of field, $F$, and electron temperature, $T_\mathrm{el}$, is given by \begin{equation}
J(F, T_\mathrm{el}) = e \int D(E_x, F)\, N(E_x, T_\mathrm{el})\,dE_x,
\label{eq:current}
\end{equation}
where $e$ is the elementary charge, $E_x$ is the energy of the electron in the surface normal direction, $D(E_x, F)$ is the transmission coefficient of the tunnel barrier, and $N(E_x, T_\mathrm{el})$ gives the distribution of electrons with normal energy $E_x$ at a given temperature.  The problem is analyzed with $x$ as the surface normal direction.  The potential inside the metal is approximated as flat, and for $x>0$, the potential energy barrier seen by the electron is simply
\begin{equation}
U(x) = \phi + \mu  - e F\, x - \frac{e^2}{16 \pi \epsilon_0 x}
\end{equation}
referenced to zero inside the metal, with $\phi$ the work function of the metal, and $\mu$ the fermi energy.  The final term is due to the image potential.  Using the WKB approximation,
\begin{equation}
D(E_x, F) \approx \exp\left(-2\; \sqrt{\frac{2me}{\hbar^2}} \int_{x_-}^{x_+} \sqrt{U(x)-E_x} \,dx\right)
\end{equation}
where $x_+$ and $x_-$ are the classical turning points.  We take $D(E_x, F) \approx 1$ when $E_x$ exceeds the barrier height.  For a free electron metal,
\begin{equation}
N(E_x, T_\mathrm{el})\,dE_x = \frac{4\pi mk_BT_\mathrm{el}}{h^3}\ln(1+e^{(\mu-E_x)/k_BT_\mathrm{el}})\,dE_x,
\end{equation}
with $h$ the Planck constant and $k_B$ the Boltzmann constant.  This simple model for tunneling is extremely successful in analyzing the current vs. voltage ($I$-$V$) characteristics of field emitters at room temperature. \cite{Gomer61}  At zero temperature, the integral can be carried out analytically to obtain the Fowler-Nordheim equation:
\begin{equation}
J = \frac{e^3 F^2}{8 \pi h \phi t^2(w)} \exp\left[-\frac{8\pi \sqrt{2 m} \phi^{3/2}}{3heF}v(w)\right].
\label{fowlernordheim}
\end{equation}
Here, $w = e^{3/2}\sqrt{F/4\pi\epsilon_0}/\phi$, and $t(w)$ and $v(w)$ are relatively slowly varying functions of order 1, related to the elliptic integrals that occur in Eq. (\ref{eq:current}). \cite{Gomer61}  Using this in conjunction with Eq. (\ref{fowlernordheim}) multiplied by the emitting area, one obtains a relationship between the measured quantities $I$ and $V$. Fig. \ref{figure:tfe1}(b) shows a Fowler-Nordheim equation fit to room-temperature $I$-$V$ data from our tip (no laser illumination) along with the corresponding family of $I$-$V$ curves ($I_{\mathrm{TFE}}(T, F)$) at elevated temperatures, which have been obtained by numerically integrating Eq. (\ref{eq:current}).  Increasing the temperature populates states above the Fermi level which enjoy an exponential increase in the tunneling transmission, resulting in enhanced emission that can be very sensitive to temperature as well as field. We refer to emission at elevated temperatures (so that the Fowler-Nordheim equation is no longer a good description) as thermally-enhanced field emission (TFE).  Note that, as we use it, this term includes thermionic emission. \cite{Murphy56}  For the parameters of our experiment, the temperature nonlinearity can scale locally as steeply as $T^{12}$.

\subsection{Ultrafast laser-induced emission}
With a mode-locked laser, it is possible to achieve both high peak intensities and few cycle pulse durations.  This light, incident upon a biased field emission tip, leads to enhanced electron emission that can be described by the processes in Fig. \ref{emission_processes}(b-d).  Proper management of these processes should allow the temporal profile of the pulse train, or even of each optical cycle, to be transferred to the temporal profile of electron emission.
\subsubsection{Multiphoton emission}
For visible light, the energy of a single photon is typically less than the work function of the metal.  Absorption of multiple photons can liberate electrons over the barrier (Fig. \ref{emission_processes}(c)).  $N$-photon emission is described by
\begin{align}
J_N = &A_N I^N [(N\hbar \omega-\phi)^2 + \frac{\pi^2}{3}(k_B T_\mathrm{el})^2 \notag\\
&-2(k_B T_\mathrm{el})^2\sum_{n=1}^\infty \frac{(-1)^{n+1}}{n^2} e^{-n(N\hbar \omega-\phi)/k_B T_\mathrm{el}}],\label{mpeq}
\end{align}
with $I$, laser intensity, $\omega$, angular frequency of the laser, and $A_N$ a constant. \cite{Girardeau-Montaut95}  Multiphoton emission has been observed in gold \cite{Ropers07} and tungsten \cite{Schenk10} tips, with $N = 4$ and $N = 3$, respectively (both experiments used a Ti:sapphire laser).  When a bias field is applied to the tip, the height of the potential barrier is reduced due to the image charge when an electron leaves the metal (Schottky effect).\cite{Gomer61}  To account for this effect, we make the substitution $\phi \rightarrow \phi - \sqrt{eF/(4\pi\epsilon_0)}$ in Eq. \ref{mpeq}.
\subsubsection{Photo-assisted field emission}
Even if $N \hbar \omega < \phi$, it is possible for electrons to tunnel through the potential barrier if a DC bias is applied.  The emitted current is described by the Fowler-Nordheim equation with an effective work function $\phi_{\mathrm{eff}} = \phi -N \hbar \omega$.  This process is called photo-assisted field emission (Fig. \ref{emission_processes}(b)), and has been observed for laser-illuminated tungsten tips both with continuous illumination \cite{Lee73} and for illumination with a femtosecond laser. \cite{Hommelhoff06a}  Since the excited electron states in metals have short lifetimes, the emission process is expected to be prompt, i.e.~should follow the laser intensity profile to first approximation, as should multi-photon emission.
\subsubsection{Above-threshold photoemission and optical field emission}
As pulse fluence increases, above-threshold photoemission ($N$-photon emission with $N$ greater than the minimum number of photons required to escape the metal) can occur (Fig. \ref{emission_processes}(d)).  This process is analogous to above-threshold ionization of atoms. \cite{Mainfray91}  For sufficiently high laser intensities, the lower photon orders may be suppressed due to the quiver energy of the electron in the laser field.  This strong-field effect has recently been observed for ultrafast laser-induced emission from gold \cite{Bormann10} and tungsten tips. \cite{Schenk10}

For sufficiently large peak intensities, the description of the process becomes particularly simple: the laser field, $F_{\mathrm{laser}}$ is strong enough that electrons can tunnel from the Fermi level during part of the optical cycle, and one can substitute $F \rightarrow F + F_{\mathrm{laser}}$ in the Fowler-Nordheim equation to estimate the laser-triggered current.  This picture is applicable when the Keldysh (or adiabaticity) parameter, $\gamma$, is less than one. \cite{Brabec00}  This parameter is defined as
\begin{equation}
\gamma = \omega \frac{\sqrt{2 m \phi}}{e F_{\mathrm{laser}}},
\end{equation}
where $m$, $e$ are the mass and charge of the electron, respectively.  The photo-field and multiphoton emission regimes occur for $\gamma\gg1$.  The quasi-static or $\gamma\ll1$ regime is interesting from the perspective of ultrafast electron sources because, since tunneling can only happen during one half of the optical cycle and is extremely nonlinear in field, electron pulses that are shorter than the optical cycle of the laser may be generated.  Experiments with tungsten tips have reached $\gamma\sim1$, \cite{Hommelhoff06, Schenk10} which is possible with 10-100 mW of laser power at $\sim100$~MHz repetition rates due to optical field enhancement by the tips.  Even in this ``intermediate" regime, numerical simulation of the Schr\"{o}dinger equation predicts sub-optical cycle features in the electron emission.\cite{Hommelhoff06}  For atoms, this has been predicted analytically by an extension of the Keldysh theory \cite{Yudin01} and also observed experimentally. \cite{Uiberacker07}
\subsection{Laser-induced thermally-enhanced field emission\label{section:tfeprocess}}
Illumination of the tip by a train of ultrafast pulses heats the tip and causes the apex electron temperature to evolve periodically in time.  Therefore, thermally enhanced field emission can increase drastically at short times following laser excitation and before the electron gas equilibrates with the lattice.  It is natural to divide the thermally enhanced current into two components: one is the DC offset (DC TFE), and the other is the pulsed component (transient TFE).
\subsubsection{Transient TFE}
The temporal profile of transient TFE depends on the electron-phonon coupling time, $\tau_\text{el-ph}$, and the dynamics of heat conduction in the tip.
The electron gas is out of equilibrium with the lattice at timescales shorter than $\tau_\text{el-ph}$, which is typically on the order of picoseconds.  Since the electron contribution to the heat capacity is smaller than that of the lattice, for pulses shorter than $\tau_\text{el-ph}$ the electrons temporarily reach high temperatures. The high electron temperature is established on a very short (fs) timescale, and its dynamics can then be described by the two-temperature model.  \cite{Anisimov74}  The electron temperature decays by thermal conduction away from the apex and by coupling to the phonons; whichever timescale is faster will determine the duration of the TFE transient.
\subsubsection{DC TFE}
Following the transient, the temperature of the tip apex is nearly constant for most of the time between laser pulses.  A temperature offset can arise from the inability of the nanostructure to dissipate the deposited energy quickly enough.  The temperature rise can be estimated as that due to a continuous laser and is approximately given by \cite{Lee89}
\begin{equation}
\Delta T \approx \frac{\sqrt{2}\beta P_0}{\pi^{3/2} \kappa w_0 \theta},\label{eq:temp_rise}
\end{equation}
where $\beta$ is the fraction of light incident on the tip that is absorbed, $P_0$ is the average laser power, $\kappa$ is the thermal conductivity of the tip material, $w_0$ is the beam waist, and $\theta$ is the opening half-angle of the tip, which is conical approaching the apex.  From $\Delta T$, the average DC TFE current can be estimated from Eq. \ref{eq:current}.
\section{Results and Discussion\label{section:results_discussion}}
\subsection{Two-photon emission vs. laser-induced thermally-enhanced field emission\label{section:two_photon_and_TFE}}
 In HfC, the laser-induced emission process exhibits two regimes; in this Section we present experimental evidence that two-photon emission dominates the current for low optical power and bias field, whereas a DC current from thermal-assisted field emission (DC TFE) dominates at high optical power and bias field.  In the range of our experimental parameters, the possibility of emission on a picosecond time-scale due to transient heating of the electron gas (transient TFE) is inconsistent with power and field dependence measurements as well as simple pump-probe measurements.

\begin{figure}
\includegraphics[width=8.6cm]{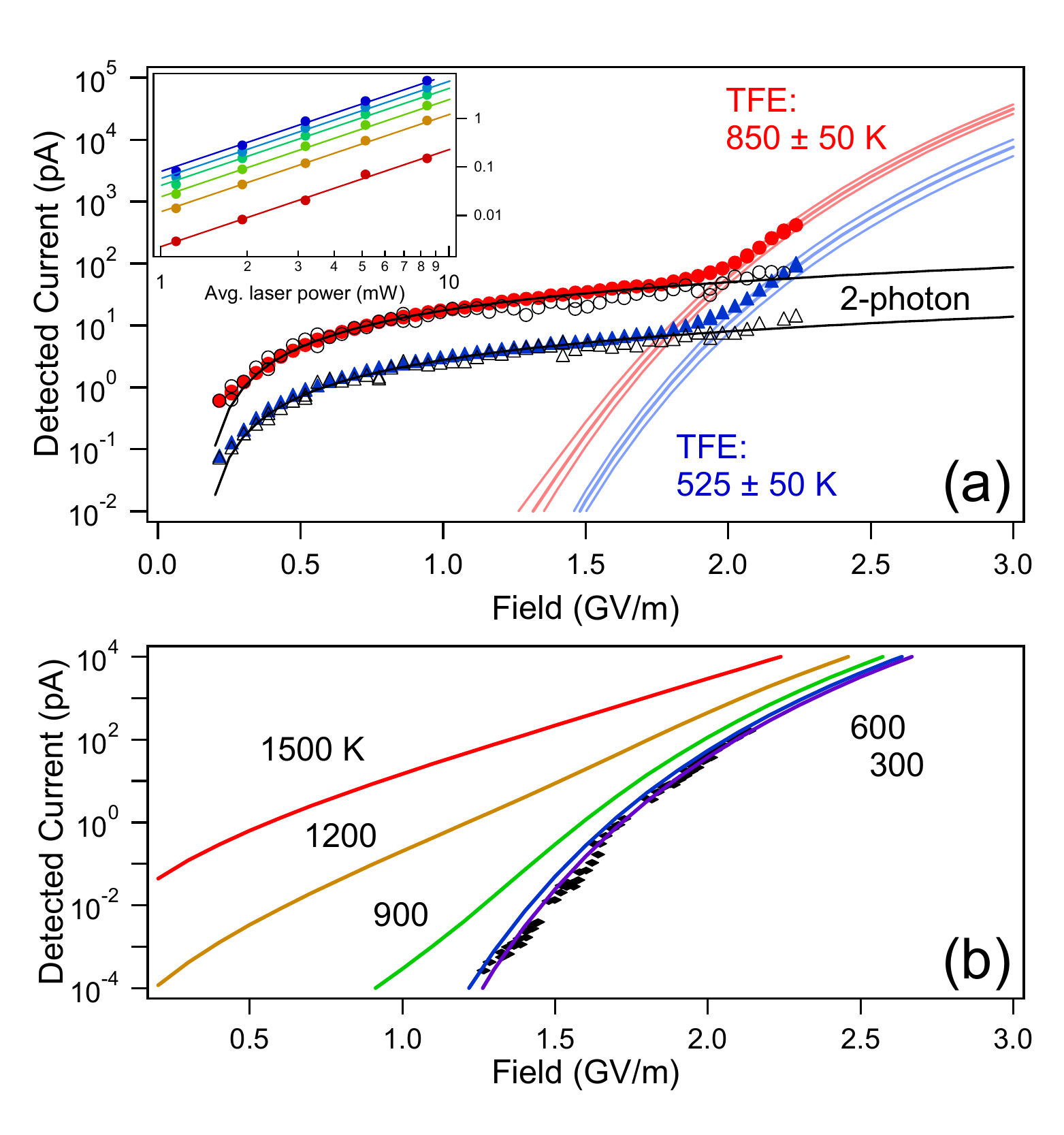}%
\caption{Ultrafast laser-triggered emission from a HfC field emission tip.  (a) Closed symbols are $I_{\mathrm{laser}}$ and open symbols are $I_{\mathrm{2f}}$, for $P = 3.9$ mW (triangles) and 9~mW (circles) laser power.  The corresponding peak intensity in the focus, neglecting field enhancement, is $2.1\times10^{10}$ and $4.9 \times 10^{10}$~W/cm$^2$.  Light solid lines are theory for TFE and dark solid lines are fits of Eq. (\ref{mpeq2}), describing two-photon emission, to the data.  Inset: Current vs. average laser power for tip bias of 200-700 V, increasing in 100 V steps.  (b) Cold field emission ($I_{\mathrm{DC}}$ at 300~K) without the incident laser.  Theory for 300 K emission is fit to the data to determine the work function and emission area.  Then the family of TFE curves (solid lines, theory) describe emission for a tip with elevated temperature, obtained using Eq. \ref{eq:current}.  To replicate the lock-in signal expected if total current is dominated by DC TFE at a given temperature, the 300 K curve is differenced from the TFE curve, to give the light solid curves in part (a) of the figure.  \label{figure:tfe1}}
\end{figure}

First, the emission regimes are demonstrated as either pulsed (at low bias) or DC (at high bias) by comparing $I_{\mathrm{laser}}$ with $I_{\mathrm{2f}}$ as the bias field is varied, as shown in Fig. \ref{figure:tfe1}(a).  The crossover between these regimes occurs around 1.8~GV/m; above this bias field,  $I_{\mathrm{2f}}$ accounts for less than half of $I_{\mathrm{laser}}$.

In the low-bias regime, we attribute the pulsed current to $N$-photon emision over the potential barrier, with $N = 2$ the dominant contribution.  Evidence for two-photon emission includes both the bias field and intensity dependence of $I_{\mathrm{2f}}$ (and, equivalently in this regime, $I_{\mathrm{laser}}$) which are plotted in Fig. \ref{figure:tfe1}(a) and the inset.  The inset shows a quadratic intensity dependence over the full range of bias fields applied, indicating a two-photon process.  Fig. \ref{figure:tfe1}(a) plots the field dependence for only two specific intensities for clarity.  The $I_{\mathrm{2f}}$ data are fit, using a single normalization coefficient, to the model
\begin{equation}
I_{\mathrm{2-ph}} \propto P^N[(N\hbar \omega - \phi+\sqrt{eF/(4\pi\epsilon_0)})^2 + \frac{\pi^2}{3}(k_BT)^2],\label{mpeq2}
\end{equation}
derived from Eq. \ref{mpeq} in the limit that $k_B T \ll (N \hbar \omega - \phi)$.  To determine the work function and the factor $k$ relating voltage to bias field, the room-temperature $I$-$V$ data (no laser illumination) is first fit to Eq. \ref{fowlernordheim} as shown in Fig. \ref{figure:tfe1}(b).  The temperature is determined from $I_{\mathrm{laser}}$ as described in the next paragraph.  Then, the two fitting parameters of Eq. \ref{mpeq2} are $N$ and $\omega$.  Choosing $N=2$ results in $\hbar \omega = 1.46$~eV, close to the center frequency of the (broad) laser spectrum.  Larger values of $N$ would lead to photon energies far outside the spectrum of the laser.

Next, we attribute the high-bias regime dominated by DC current to the steady state TFE discussed in section \ref{section:tfeprocess}, caused by laser heating and the tip's inability to cool to room temperature between incident optical pulses.  Above 1.8~GV/m, $I_{\mathrm{laser}}$ is well-described by fitting to temperature-dependent field emission theory (Section \ref{section:fieldemission}) as shown in Fig. \ref{figure:tfe1}(a).  The fitting is performed as follows: first, the family of TFE curves $I_{\mathrm{TFE}}(T, F)$ is calibrated as in Fig. \ref{figure:tfe1}(b) by fitting room-temperature $I$-$V$ data to the temperature-dependent field emission theory described in Section \ref{section:fieldemission}.  The curves have been obtained by numerically integrating Eq. \ref{eq:current}.  At a given temperature, the expected lock-in signal due to TFE is then calculated by subtracting room-temperature current from elevated-temperature current.  The predicted lock-in signal for different assumed tip apex temperatures is shown as the light solid lines in Fig. \ref{figure:tfe1}(a).  The apex temperatures in the experiment may be inferred to $\pm$ 50~K with this method, and agree with the prediction of Eq. \ref{eq:temp_rise} for our experimental parameters, as will be discussed in Section~\ref{section:modeling}.

\begin{figure}
\includegraphics[width=8.6cm]{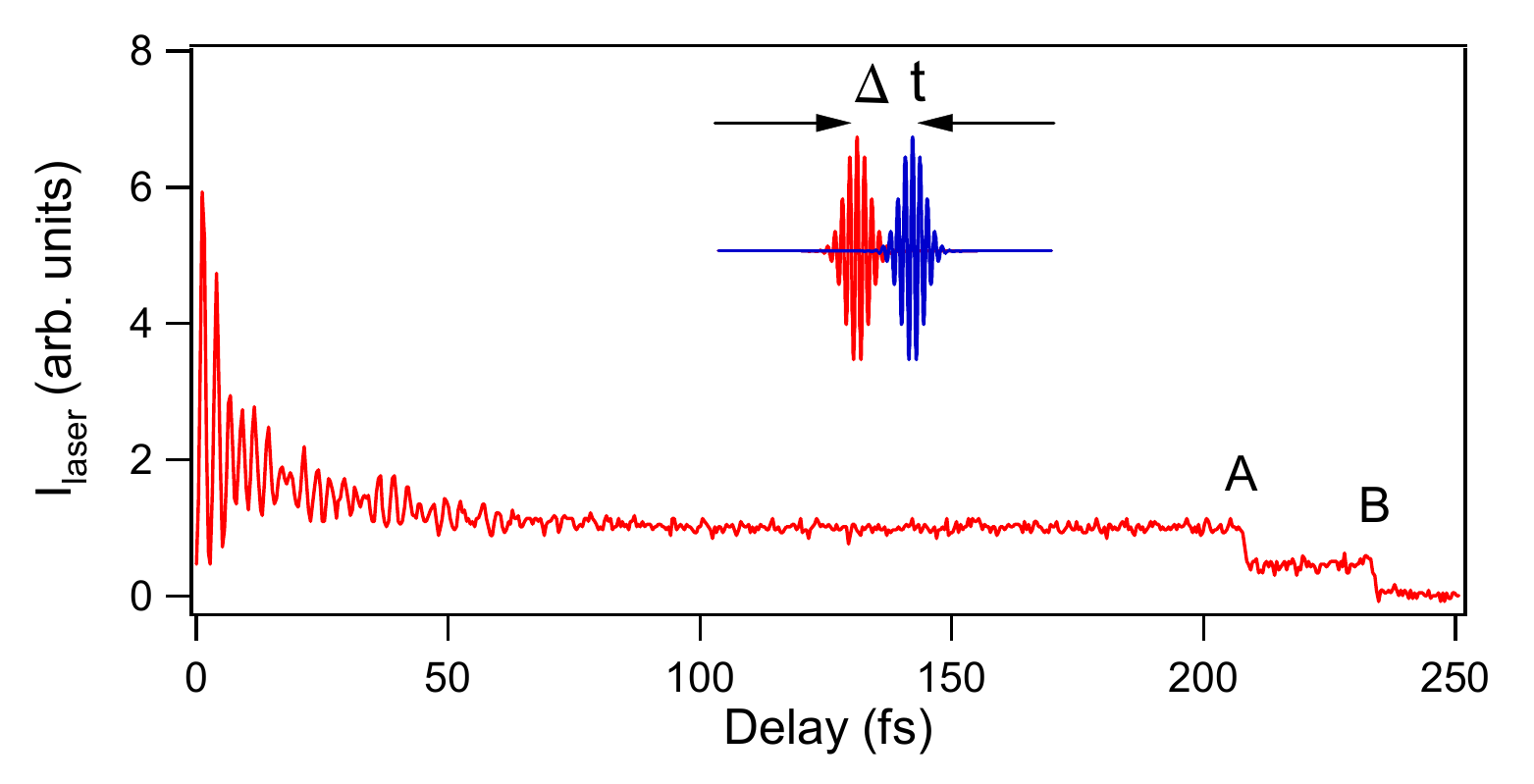}%
\caption{$I_{\mathrm{laser}}$ as a function of time delay between two identical pulse trains incident on the tip.  For $\Delta t \gtrsim50$~fs (well away from interference between the two pulses), $I_{\mathrm{laser}}$ is independent of the time delay.  At $A$, one of the two pulse trains is blocked, showing that the signal decreases to half of its value (within $\sim5\%$, due to a small power imbalance or difference in alignment between the two pulse trains).  At $B$, both beams are blocked.  The average power is 4 mW per pulse train, and the tip bias field is 0.4 GV/m. In this parameter regime, $I_{\mathrm{2f}}$ and $I_{\mathrm{laser}}$ are equivalent.\label{figure:ac}}
\end{figure}

One may consider whether transient TFE (due to emission from the hot electron distribution on a picosecond timescale after the laser pulse, as described in section \ref{section:tfeprocess}) contributes significantly to current in the pulsed regime.  In that case, the emission timescale could be on the order of the electron-phonon coupling time ($\sim$1~ps), significantly longer than the $\sim$10~fs expected for two-photon emission.  However, transient TFE can be excluded as the dominant current source for several reasons.  First, we observe an additive baseline for autocorrelation traces measured with the tip as a detector \cite{Hommelhoff06}, as shown in Fig. \ref{figure:ac}.  This is seen for time delays $\Delta t \gtrsim50$~fs, to avoid fringes from the interference of the two pulse trains at the tip.  Transient TFE should not exhibit such linearity when the total energy deposited is doubled.  Secondly, the expected field and power dependence of transient TFE is qualitatively different from that of two-photon emission.  Current due to transient TFE would scale roughly like the TFE curve (Fig. \ref{figure:tfe1}(b)) for the corresponding peak electron temperature following the laser pulse, up to a normalization factor due to the finite electron-phonon coupling timescale.  For these curves, the nonlinearity of TFE current as a function of laser power changes with bias field.  This is in stark contrast to the observed $P^2$ behavior of $I_{\mathrm{2f}}$, independent of bias field, and can be immediately visualized by noting that whereas two-photon emission curves vs. field are parallel for various laser powers on a semi-logarithmic plot (Fig. \ref{figure:tfe1}(a)), the family of TFE curves (Fig. \ref{figure:tfe1}(b)) are not.

\begin{figure}
\includegraphics[width=8.6cm]{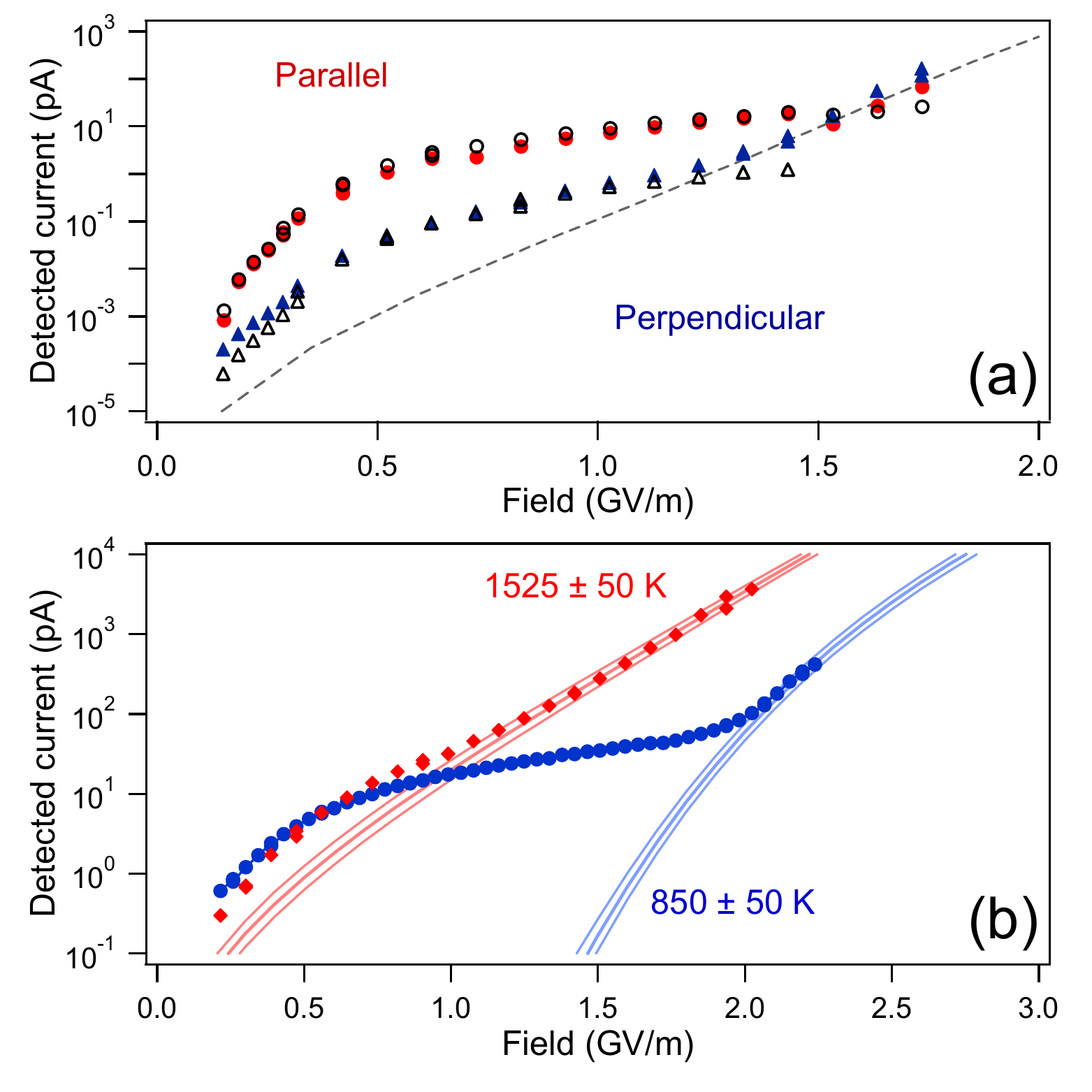}%
\caption{(a) Polarization strongly influences the two-photon emission.  Circles (triangles) are for polarization parallel (perpendicular) to the tip axis.  Closed (open) symbols are $I_{\mathrm{laser}}$ ($I_{\mathrm{2f}}$).  The dashed line is the calculated TFE current for the tip at 1100~K.  The ratio of $I_{\mathrm{2f}}$ for the two polarization cases is constant. (b) Misalignment can excessively heat the tip.  Circles (blue) are $I_{\mathrm{laser}}$ with the laser aligned to the apex of the tip, whereas diamonds (red) are $I_{\mathrm{laser}}$ measured with the laser aligned further up the shank, for the same laser power.  Light solid lines are DC TFE theory at the labeled temperatures.  Alignment up the shank increases the TFE current due to the higher apex temperature, but decreases the two-photon emission, observable at low fields.  The average laser power is 9 mW for all of the curves (peak intensity in the focus, $4.9\times 10^{10}$ W/cm$^2$)\label{tfe2}}
\end{figure}

The effect of laser polarization also supports two-photon emission in favor of transient TFE.  For laser polarization perpendicular to the tip axis, $I_{\mathrm{2f}}$ is suppressed by a factor of $\sim$25 independent of the bias (Fig. \ref{tfe2}(a)) relative to parallel polarization.  This suppression is consistent with the idea that photons excite electrons in surface states, where momentum conservation favors absorption of light with polarization vector normal to the surface \cite{Venus83}.  Light polarized parallel to the tip axis satisfies this condition near the apex, where both DC and optical field enhancement are largest.  As will be discussed in Section~\ref{section:modeling}, the transient temperature rise is expected to be larger for parallel polarization, so transient TFE would also be suppressed by perpendicular polarization.  However, the measured suppression is a constant factor independent of bias, which is again inconsistent with thermal emission at two different temperatures.

We also note that the laser heating leading to TFE in the high-bias regime is greatest when the beam is aligned slightly up the shank \cite{Lee80} because more energy is absorbed.  In Fig.~\ref{tfe2}(b), such an alignment change ($\sim$2-3~\textmu m) results in a much higher temperature.  Because the two-photon emission (the dominant contribution to $I_{\mathrm{laser}}$ for low bias fields) depends mostly on the intensity of light at the apex, it decreases when the beam is aligned up the shank and the two curves in Fig.~\ref{tfe2}(b) cross each other near 0.5 GV/m.  All other data presented in this work is measured with the beam aligned to the apex by maximizing $I_{\mathrm{2f}}$.

\begin{figure}
\includegraphics[width=8.6cm]{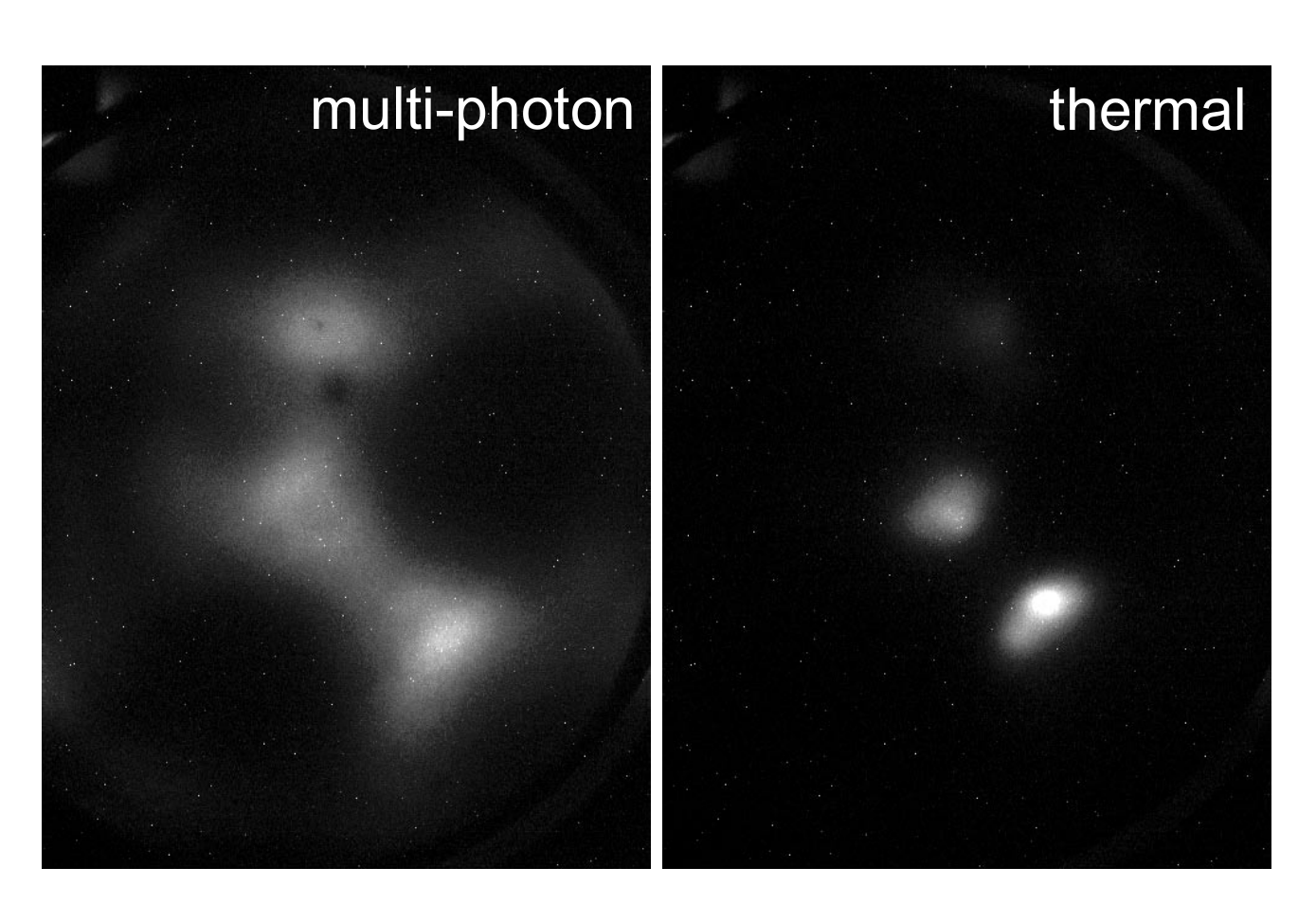}%
\caption{Emission pattern under laser illumination for two different DC bias fields applied to the tip.  At lower bias fields (left), multi-photon emission dominates while at higher bias fields (right) thermally-enhanced field emission takes over.  Cold field emission with no illumination gives a pattern identical to the TFE whereas the multiphoton emission is more diffuse and emits from slightly different crystal planes. \label{fem}}
\end{figure}

Besides the identification of the two emission processes via $I$-$V$ measurements, we note that the emission pattern is significantly different for the two processes (Fig. \ref{fem}).  When TFE dominates, the pattern is essentially identical to the DC emission pattern without laser illumination.  However, when two-photon emission dominates, the pattern is more diffuse and comes from different crystal planes.  We have observed this for each of several different 310-oriented HfC tips, allowing for the rapid identification of the dominant emission mechanism.  We note that this differs from the case of photofield emission from 111-oriented tungsten \cite{Hommelhoff06a}, where the prompt laser-induced emission has the same emission pattern as DC emission.

\subsection{Model of laser heating and temperature dynamics}\label{section:modeling}
\begin{figure}
\includegraphics[width=8.6cm]{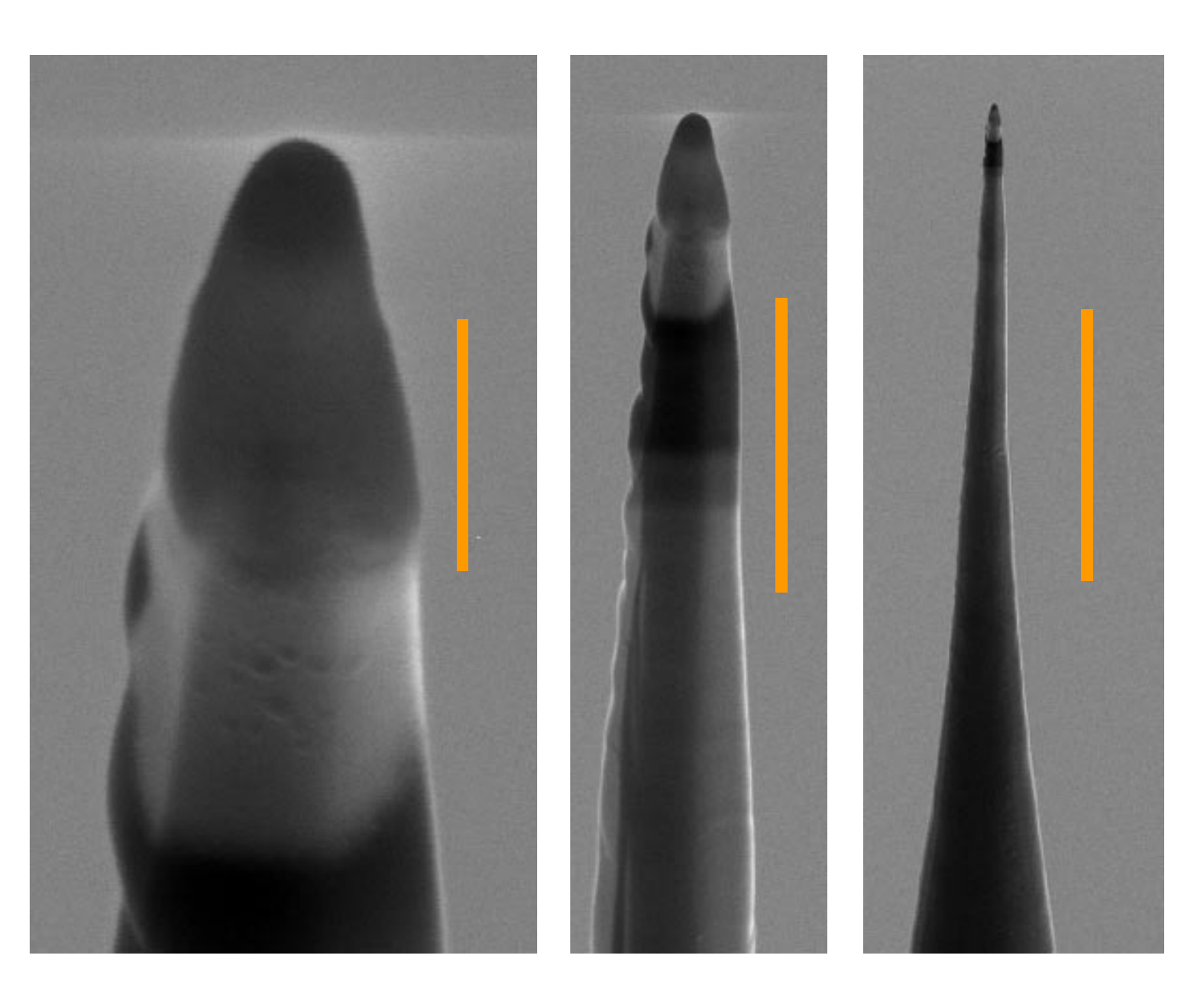}%
\caption{Images of the HfC tip taken with a scanning electron microscope.  The scale bars, from left to right, are 0.5, 2, and 10~\textmu m.  The viewing perspective is such that the tip is directed out of the plane by $\sim45\deg$.\label{sem}}
\end{figure}
A finite element analysis model\cite{comsol3.5a} of the evolution of the temperature at the tip apex following laser illumination is developed.  The model allows us to predict how transient TFE scales with laser power and bias field in comparison with the prompt two-photon emission of interest (see Sec. \ref{section:transient_bounds}).  The first stage of the model is a solution for the steady-state temperature distribution in the tip under conditions of equilibrium between the electron gas and the lattice.  This is used as the initial state of the tip before the arrival of the next laser pulse in the second stage of the model.  The second stage solves for the evolution of $T_\mathrm{el}$ in the first ns after the arrival of the laser pulse.  Representative material parameters used in the models are given in Table \ref{Table:MaterialParameters}.
\begin{table}[tb]
\caption{\label{tab:modeling_params}Materials properties used for modeling.\label{Table:MaterialParameters}}
\begin{ruledtabular}
\begin{tabular}{lcccc}
                                              & T~(K) &  HfC   &   W   &  Au   \\ \cline{2-5}
\multirow{2}{*}{$C$~($10^6$~J/(m$^3$ K))}\cite{Pierson96, NISTChemWebBook, Takahashi86}     &  300  &  2.29  &  2.55 &  2.48\\
                                              & 1000  &  3.48 &  2.89 &  2.88\\ \cline{2-5}
\multirow{2}{*}{$C_\mathrm{el}$~($10^4$ J/(m$^3$ K))}\cite{Ihara76, Lin08}  &  300  &  3.81\footnote{HfC $C_\mathrm{el}(T)$ determined from the density of states in Ref. \onlinecite{Ihara76}.}   &  3.46  &  2.02  \\
                                             & 1000  &  8.42\footnotemark[\value{mpfootnote}] & 9.88   &  6.73 \\\cline{2-5}
\multirow{2}{*}{$\kappa$~(W/(m K))}\cite{Wuchina04}$^,$\cite{CRC92}                &  300  &  20.7  &  171  &  317  \\
                                              & 1000  &  27.0 &  117 &  270  \\ \cline{2-5}
$\kappa_\text{el,0}$~(W/(m K))\cite{CRC92}$^,$\cite{Christensen07}    &    &  ---\footnote{For HfC, $\kappa_\mathrm{el}$ and $\kappa_\mathrm{ph}$ were taken to be 12 W/(m K), half of $\kappa$ at 630 K, the approximate initial temperature (see text).}    &  136\footnote{W $\kappa_\mathrm{el}$ at 400 K determined by fitting electrical resistivity and thermal conductivity data in the 300-900 K range (Ref. \onlinecite{CRC92}) using the Wiedemann-Franz law.}   &  318  \\ \cline{3-5}\noalign{\vspace{0.2mm}}
$g_\mathrm{el-ph}$ ($10^{16}$ W/(m$^3$ K))\cite{Lin08}     &			  &		6.85
\footnote{Example value used in HfC TTM model.}  &	19.1 &  2.61\\ \cline{3-5}\noalign{\vspace{0.1mm}}
\multirow{2}{*}{$\epsilon_\mathrm{r}$}\cite{Fluck82}$^,$\cite{CRC92}$^,$\cite{Johnson72}                    &Re[]       &  9.49\footnote{Determined by Drude model fit to reflectivity curves in Ref. \onlinecite{Fluck82} using two Lorentz oscillator terms.}  &  5.2  & -24   \\
                     &     Im[]  & -8.48\footnotemark[\value{mpfootnote}]  & -19.4 & -1.5  \\ \cline{3-5}
$\beta$ ($\parallel$ polarization)                &       &  0.8   &  0.6  & 0.037 \\
$\beta$ ($\perp$ polarization)            &       &  0.95  &  0.7  & 0.047 \\
\end{tabular}
\end{ruledtabular}
\end{table}

\begin{figure}
\includegraphics[width=8.6cm]{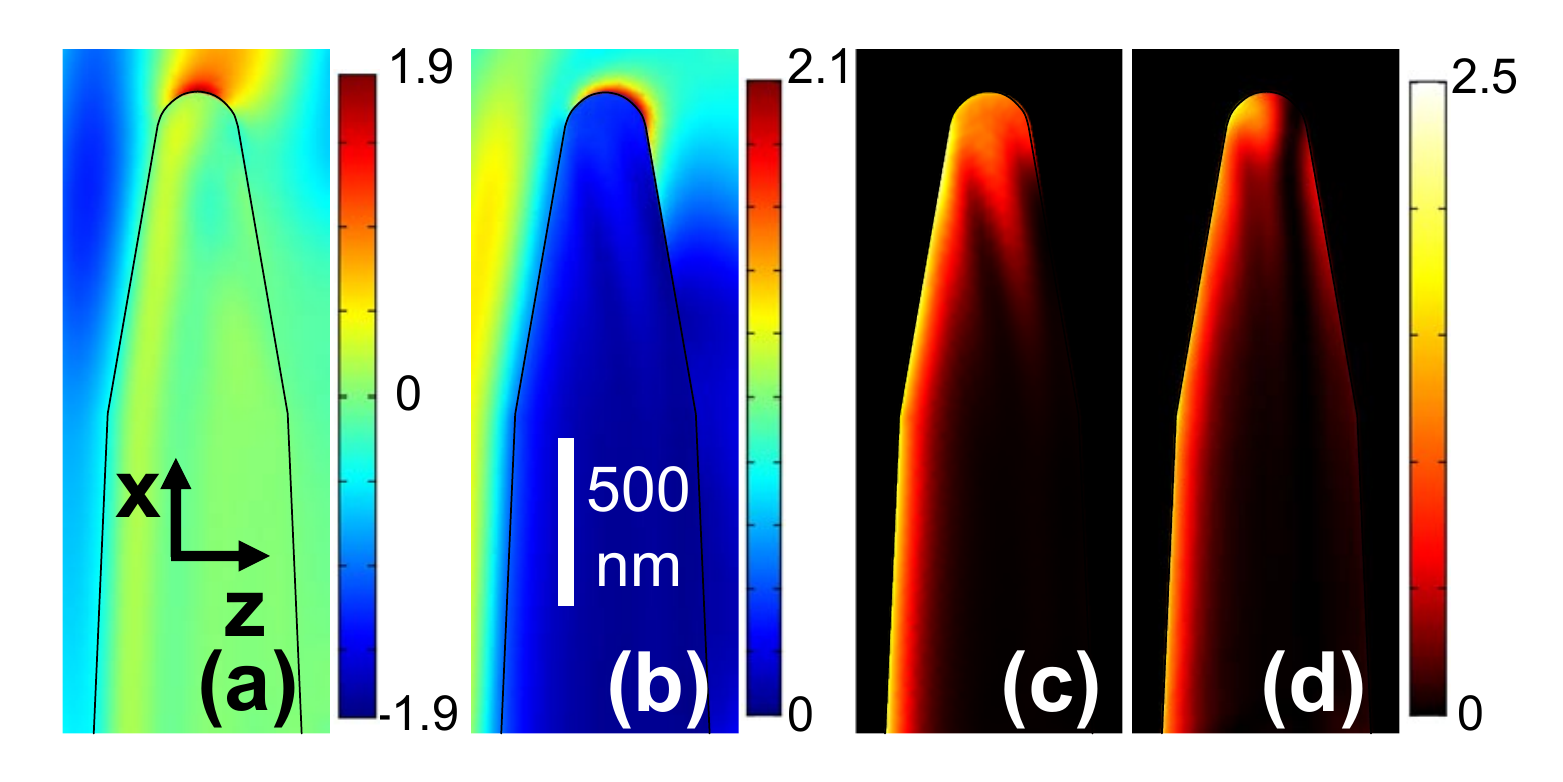}%
\caption{Modeling the laser-induced heating.  An incident 1 V/m monochromatic plane wave is assumed, and the harmonic response of the electric fields is solved with finite element analysis software. (a) $\mathbf{E_x}$ at the particular phase when it reaches its maximum for an $x$-polarized incident wave.  (b) Cycle-averaged magnitude of the electric field for incident $x$-polarized light. (c), (d) Average power density heating the tip for $x$- and $y$-polarized incident light, respectively.  Electric field units are V/m and power density units are $10^4$ W/m$^3$.  All plots are for the cross section given by the symmetry plane ($y=0$) of the model.\label{ModelDetails}}
\end{figure}

The tip geometry is based on SEM characterization of the tip used in the experiments; for the 120-nm radius tip imaged in Fig. \ref{sem}, the upper 1.8 microns of the model are shown in Fig. \ref{ModelDetails}.  The $2.5$~degree half-angle cone extends to 16~\textmu m below the apex before slowly flaring out to a 500~\textmu m diameter cylinder held at 300~K.  A hybrid model was used for the spatial distribution of heating in the tip in order to make feasible the computation of heating near the apex, where optical field enhancement \cite{Martin01, Yanagisawa10a} has complicated effects.  Maxwell's equations are solved near the tip apex, while a simple calculation of the energy intersected by the tip's silhouette is used as the heat source further up the shank, as described below.

Fig. \ref{ModelDetails} shows how the detailed heating distribution is modeled near the tip apex for an incident 800-nm plane wave.  Fig. \ref{ModelDetails}(a) depicts the axial component of the electric field at the phase when it reaches its maximum strength (field enhancement 1.9), and Fig. \ref{ModelDetails}(b) shows the cycle-averaged amplitude of the electric field, both for $x$-polarized incident light.  The plane wave of amplitude 1 V/m is incident from the left side of the figure, and the highest time-averaged fields occur toward the shadow side of the apex in agreement with Ref.~\onlinecite{Yanagisawa10a}.  The accuracy of the model was checked by replicating the models in Ref.~\onlinecite{Yanagisawa10a} and by benchmarking against Mie scattering calculations for a 100-nm gold sphere.

The field solutions are used to calculate the power density of resistive heating inside the tip, $p_{\mathrm{model}}(x,y,z)$, as shown in Fig. \ref{ModelDetails}(c,d) for $x$- and $y$-polarized incident light, respectively.  We approximate the resistive heating caused by the broad spectrum of the laser by the heating due to 800~nm light.  The integrated resistive heating changes by less than 10\% if the problem is solved for either 680~nm or 940~nm incident light, and the distribution does not noticeably change.

Although the electromagnetic model is solved using an incident plane wave, the experiment uses a focused beam.  We approximate the heat source in the apex region as
\begin{equation}
 p_{\mathrm{apex}}(x,y,z)=p_{\mathrm{model}}(x,y,z) \; e^{-2(x/w_0)^2}
 \end{equation}since the beam waist is much larger than the size scale of the tip geometry.  Further up the shank, the time-averaged volumetric heat source $p_{\mathrm{shank}}(x)$ is taken to be
\begin{equation}
p_{\mathrm{shank}}(x) = \frac{4\beta P_0}{\pi w_0^2 r(x)}e^{-2(x/w_0)^2},
\end{equation}
where $r(x)$ is the radius of the tip shank at distance $x$ from the apex and $P_0$ is the laser's average power.  The constant factor $\beta$ takes into account the interrelated effects of field enhancement and reflectivity of the tip material, and is determined by requiring that $p_{\mathrm{shank}}(x)$ smoothly match the numerical solution (integrated over $y$, $z$) near the apex.  Although the expression for $p_{\mathrm{shank}}$ ignores the finite penetration depth of heating due to skin effects, details of the shank heating distribution affect neither the equilibrium temperature distribution nor the short-timescale dynamics of the temperature near the apex, which are the quantities of interest.

To model the steady-state temperature rise of the tip during laser heating, we use a continuous (i.e. not pulsed) heat source,
 \begin{equation}
 p(x, y, z) =\begin{cases}p_{\mathrm{apex}}(x, y, z),& x<1.5\text{ \textmu m}\\
 p_{\mathrm{shank}}(x),& x>1.5\text{ \textmu m}
 \end{cases}
\end{equation}
for $t>0$.  Temperature-dependent values of the thermal conductivity and heat capacity of HfC were taken from the literature. \cite{Wuchina04, Pierson96}  The model predicts a DC temperature rise of 262~K and 567~K for average laser power of 3.9 mW and 9 mW, respectively.  These values are in good agreement with the observed values of 225$\pm$50~K and 550$\pm$50~K (see Fig. \ref{figure:tfe1}).  The simple model of Eq. \ref{eq:temp_rise} predicts apex temperature increases of 300~K and 700~K for 3.9 and 9 mW, with $\theta = $ 2.5 degrees and $\beta = 0.95$ (note that Eq. \ref{eq:temp_rise} assumes a simpler tip geometry).  The time-scale for reaching the equilibrium temperature after the heat source is turned on is tens of microseconds.
\begin{figure}
\includegraphics[width=8.6cm]{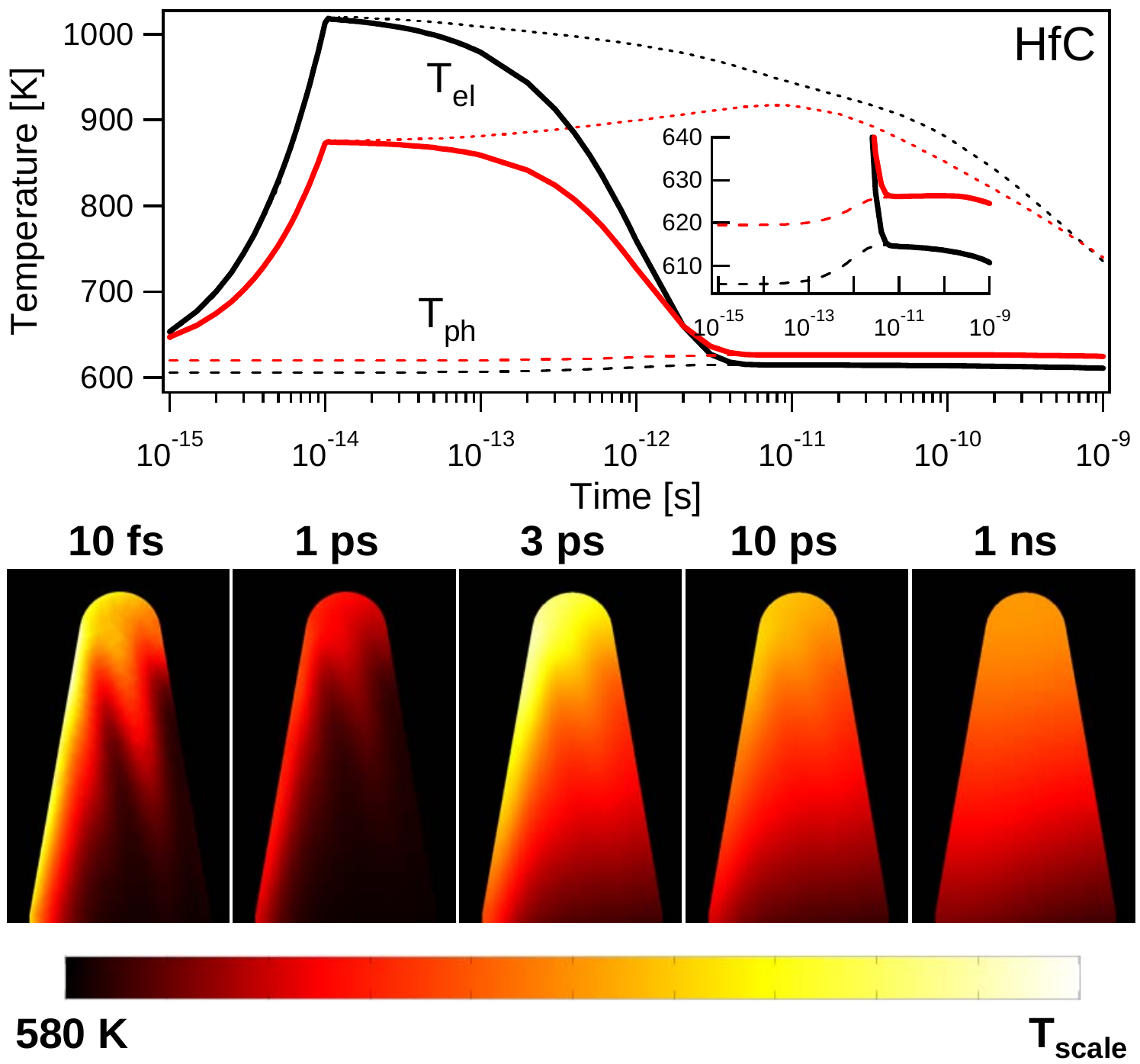}%
\caption{Results of the TTM simulation for HfC (details of the model are provided in the text).  Upper plot:  Temperature at the tip apex as a function of time.  Black (darker) and red (lighter) curves are for $x$- and $y$-polarization, respectively.  Thick solid curves are $T_\mathrm{el}$ and thin dashed curves are $T_\mathrm{ph}$.  Thin dotted curves show $T_\mathrm{el}$ without electron-phonon coupling for comparison.  The inset is rescaled to show $T_\mathrm{ph}$ more clearly.  Lower plots:  the $T_\mathrm{el}$ distribution ($y$ = 0 cross-section) at several times after a laser pulse.  For clarity $T_\mathrm{scale}$, the maximum temperature of the color scale, is 1250 K for the 10~fs and 1~ps frames, and 640~K otherwise. \label{figure:TapexDynamics}}
\end{figure}
The second stage of the model investigates the effect of a single laser pulse once the steady-state temperature profile has been established.  The electron and lattice temperatures ($T_\mathrm{el}$ and $T_\mathrm{ph}$) are described by the coupled equations
\begin{align}
&C_e \frac{d T_\mathrm{el}}{dt} = \nabla\cdot(\kappa_\mathrm{el} \nabla T_\mathrm{el}) - g_\text{el-ph} (T_\mathrm{el}-T_\mathrm{ph})+p(x, y, z)f(t)\\
&C_\mathrm{ph} \frac{d T_\mathrm{ph}}{dt} = \nabla\cdot(\kappa_\mathrm{ph} \nabla T_\mathrm{ph}) + g_\text{el-ph} (T_\mathrm{el}-T_\mathrm{ph}),
\end{align}
where $g_\text{el-ph}$ is the electron-phonon coupling constant.  $T_\mathrm{el}$ and $T_\mathrm{ph}$ are functions of position and time, and $C_{e,\; \mathrm{ph}}$ and $\kappa_{e,\; \mathrm{ph}}$ may be functions of position and time via their dependence on $T_\mathrm{el}$ and $T_\mathrm{ph}$.  The temporal profile of the heat source, $f(t)$, is on at a constant value for $0\leq t \leq 10$ fs and zero otherwise.  It is normalized to represent the heating from a single pulse in the 150~MHz pulse train for 4.6~mW average power.  The initial condition for $T_\mathrm{el}$ and $T_\mathrm{ph}$ is the temperature profile obtained for steady-state heating with 4.6~mW average power.  The electron thermal conductivity for HfC is unreported in the literature but is expected to be similar to that of TiC, approximately half the total thermal conductivity. \cite{Pierson96}   Since the electron-phonon coupling time is unknown, two solutions are shown in Fig. \ref{figure:TapexDynamics}: the electron temperature with no electron-phonon coupling ($g_\text{el-ph}=0$; thin dotted curves) and solutions for both $T_\mathrm{el}$ and $T_\mathrm{ph}$ (thick curves and thin dashed curves, respectively) assuming $g_\text{el-ph}=6.85 \times 10^{16}$ W/(K m$^3$), independent of temperature, which corresponds to $\tau_\text{el-ph} \sim C_\mathrm{el}/g_\text{el-ph} \sim 1$ ps.  The solution with no electron-phonon coupling is an upper limit to the electron temperature.  The apex temperature in this solution decays via conduction on a timescale of hundreds of ps; normal values of electron-phonon coupling times are on the order of picoseconds, which implies that (barring an anomalously long $\tau_\text{el-ph}$) the duration of TFE from the HfC tip following a laser pulse is set principally by the electron-phonon coupling time.  The electron temperature distribution near the tip apex for the HfC model with electron-phonon coupling is also shown in Fig. \ref{figure:TapexDynamics}.  Note the change in the temperature scale between the 1~ps and 3~ps plots as the electron temperature decreases due to electron-phonon coupling without significant spatial redistribution of thermal energy from conduction.
\begin{figure}
\includegraphics[width=8.6cm]{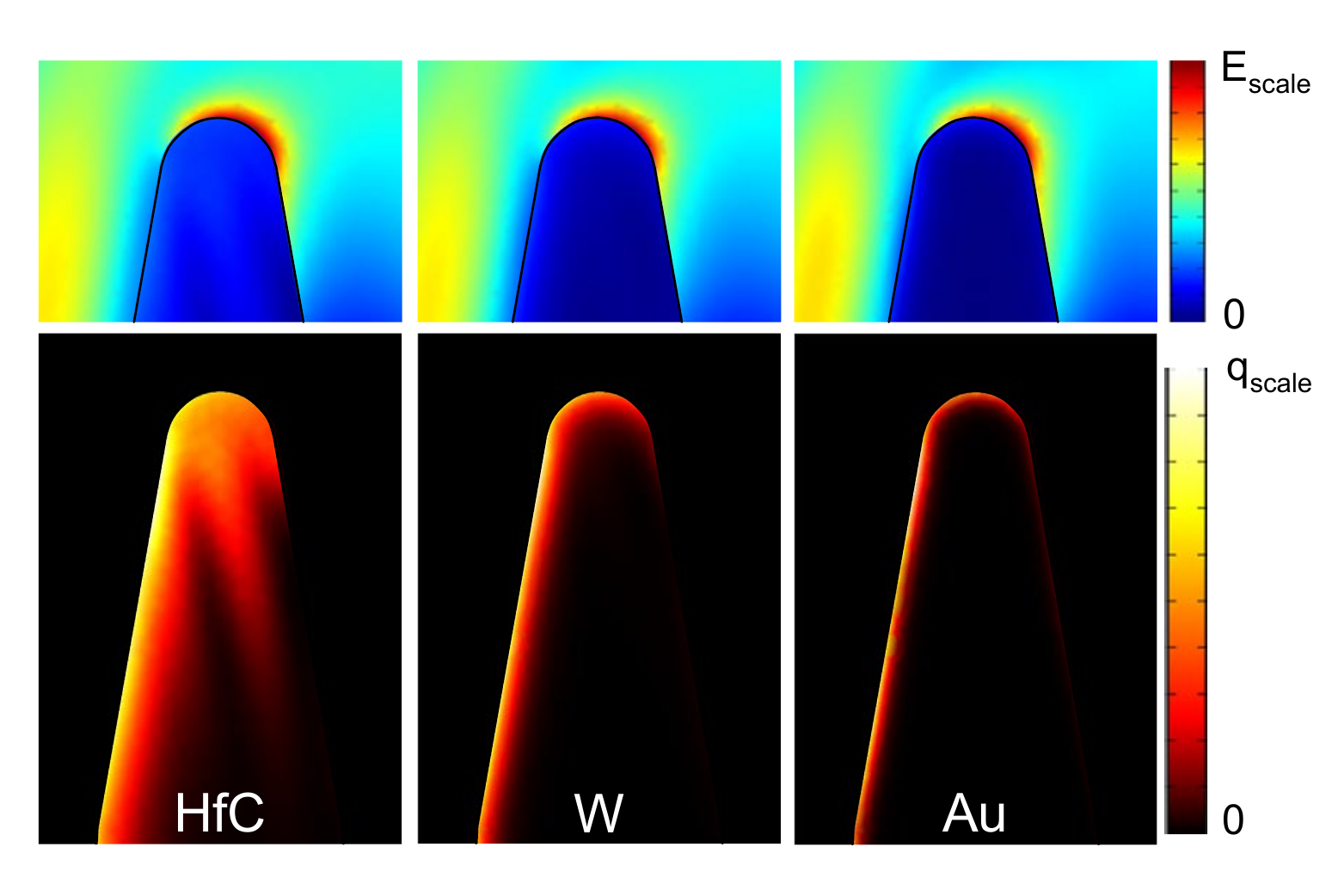}%
\caption{(Top row) Modeled, cycle-averaged electric field strength near the tip apex for a 1 V/m, 800 nm, $x$-polarized incident plane wave.  $E_{\mathrm{scale}}$ is 2.1, 2.2, and 2.4 V/m for HfC, W, and Au, respectively.  (Bottom row) Energy density deposited by a single laser pulse in the 4.6 mW, 150 MHz pulse train when focused to a 2.7~\textmu m waist.  The color scale ranges from 0 to $q_{\mathrm{scale}}$ = 5.0, 8.5, and $0.8\times10^7$~J/m$^3$ for the three materials, respectively.  \label{Enorm_q}}
\end{figure}

The model can be used to compare HfC, tungsten, and gold as tip materials.  Fig. \ref{Enorm_q} shows the time-averaged magnitude of the electric field near the apex and the heat deposited by a single pulse in the 4.6~mW pulse train, for $x$-polarized incident light.  Although the field enhancement outside the tip is only affected at the 10\% level by the various materials' permittivities, the heat density deposited at the apex varies by an order of magnitude and the total heat deposited in the gold tip is 26 times less than for HfC.  However, the smaller skin depth, but similar reflectivity, of tungsten as compared to HfC leads to a 70\% higher heat density impulsively deposited in the electron gas at the apex.
\begin{figure}
\includegraphics[width=8.6cm]{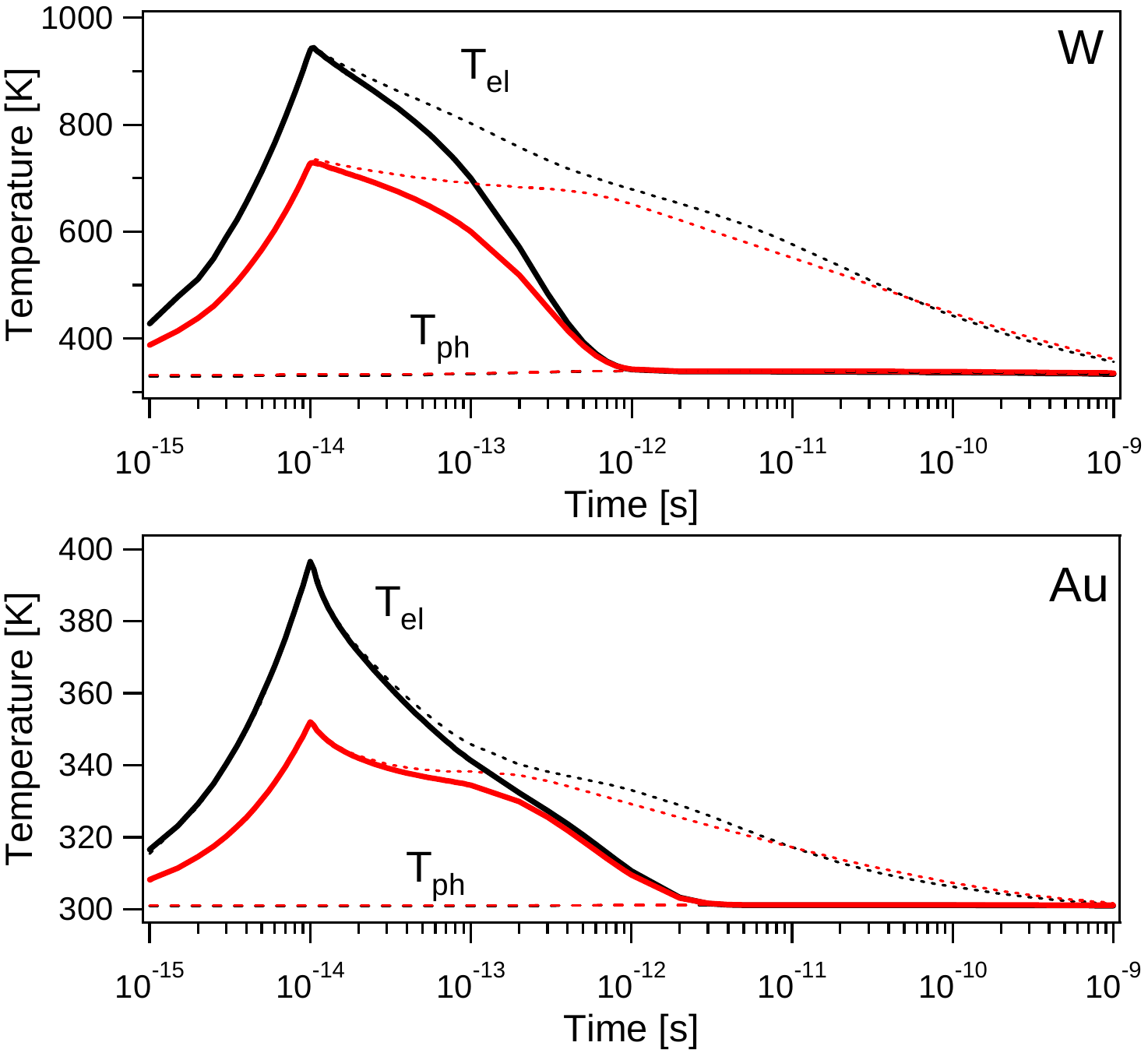}
\caption{Results of the TTM simulation for W and Au.  The curves are plotted in the same way as in Fig. \ref{figure:TapexDynamics}.\label{figure:TTM_3materials}}
\end{figure}

Fig. \ref{figure:TTM_3materials} shows the modeled apex electron temperature due to transient heating from a laser pulse for tungsten, and gold with the same experimental parameters as Fig. \ref{figure:TapexDynamics} and material parameters in Table \ref{Table:MaterialParameters}.  The electron thermal conductivity for W and Au as a function of electron and lattice temperature is expanded as\cite{Vestentoft06}
\begin{equation}
\kappa_\mathrm{el}(T_\mathrm{el}, T_\mathrm{ph}) = \kappa_{\mathrm{el, }0} \frac{T_\mathrm{el}}{T_\mathrm{ph}}.
\end{equation}
Conduction of the lattice is insignificant and therefore neglected ($\kappa_\mathrm{ph} = 0$).  The values of $g_\text{el-ph}$ (see Table \ref{Table:MaterialParameters}) are taken from Ref. \onlinecite{Lin08}, averaged over the temperature range of the simulation.  Several qualitative trends are captured in Fig. \ref{figure:TTM_3materials}.  First, due to gold's high reflectivity and both gold and tungsten's high thermal conductivities, the initial steady-state elevated apex temperature is very low compared to that of HfC.  Therefore DC TFE is essentially negligible in tungsten and gold for the range of average optical power used in high repetition rate experiments.\cite{Hommelhoff06a, Hommelhoff06, Ropers07, Barwick07, Ganter08, Yanagisawa10}  Second, $x$-polarization heats the tip apex more than $y$-polarization for all three materials, as observed by the initial temperature rise for the single pulse.  For $y$-polarization in HfC, the heat is primarily deposited slightly away from the apex (see Fig. \ref{ModelDetails}(d)), so that, without electron-phonon coupling, the apex temperature continues to rise after the optical pulse due to thermal conduction.  Third, for gold and tungsten, the apex cools via diffusion much more rapidly than HfC in the absence of electron-phonon coupling.  This is due to both the higher electronic thermal conductivities of these two materials and the fact that the heating is initially more narrowly confined (smaller optical skin depth).  In gold, because the mean free path of the excited electrons is larger than the optical skin depth, ballistic transport would reduce the apex temperature even further.\cite{Wellershoff99, Houard11}

\subsection{Bounds on transient thermally-enhanced field emission\label{section:transient_bounds}}
The fractional contribution of transient TFE to the pulsed current (measured by $I_{\mathrm{2f}}$), as well as its scaling with laser power, should be compared with that of multi-photon emission in order to understand the regimes in which the tip can be operated as a fs-timescale emitter.  We note that the field dependence of $I_{\mathrm{2f}}$ (Figs. \ref{figure:tfe1}, \ref{tfe2}) is accurately fit by two-photon emission using Eq. \ref{mpeq2} throughout the parameter regime accessible in our experiment (limited to $P\lesssim11$~mW, above which the degree of DC TFE induced by small mis-alignments of the beam is much larger than the pulsed current).  This suggests that transient TFE contributes negligibly to the total pulsed current.
\begin{figure}
\includegraphics[width=8.6cm]{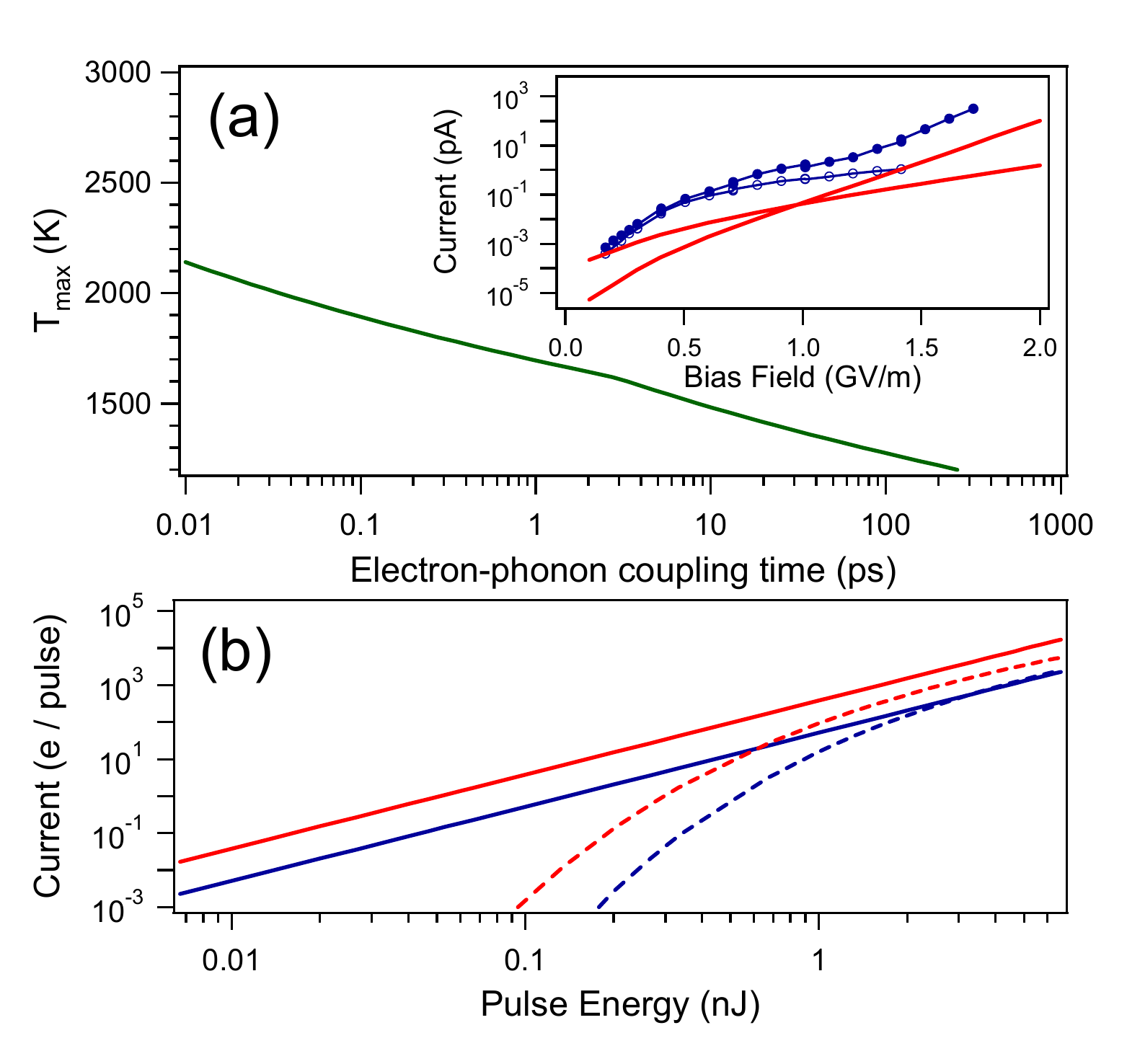}%
\caption{Upper bound to peak temperature at the tip apex immediately after heating by a single pulse in a 150~MHz pulse train of 10 mW average power focused to 2.7~\textmu m, as a function of assumed $\tau_{e-\mathrm{ph}}$.  The inset provides an example calculation: Open circles are $I_{\mathrm{2f}}$ and closed circles are $I_{\mathrm{laser}}$. The two solid curves represent transient TFE at different temperatures, scaled by $\tau_{e-\mathrm{ph}}$ so as not to exceed $I_{\mathrm{2f}}$. The steeper curve is 1200~K emission for 257~ps, and the more shallow curve is 2000~K emission for only 34~fs.  (b) Scaling of two-photon emission (solid lines) and the estimated transient TFE (dashed curves) for a single pulse incident on a room-temperature tip.  The lower (blue) and upper (red) curve in each pair are for a 0.5 and 1.5~GV/m bias field, respectively, the heat deposited per nJ pulse energy was taken from the FEM model, and 1~ps electron-phonon coupling time has been assumed.  The largest pulse energy shown corresponds to $\gamma\sim1.5$, ignoring the Schottky barrier reduction.\label{figure:trans_bound}}
\end{figure}

We can place the most stringent bound on the contribution of transient TFE using experimental data for 10~mW (the highest power measured) with perpendicularly-polarized light (suppressing two-photon emission by a factor of $\sim30$).  This requires an estimate of the field dependence of the transient TFE, as follows:  First, the measured DC thermal emission at room temperature is used to calibrate the family of TFE curves $I_{\mathrm{TFE}}(T, F)$ for this particular data set, as in Fig. \ref{figure:tfe1}(b).  Next, a conservative upper limit to the transient TFE as a function of bias field is estimated as a square current pulse lasting for a time $\tau_\text{el-ph}$, due to the peak electron temperature before coupling with the phonons.  Therefore, the upper limit to transient TFE is $I_{\mathrm{transient}}(F)<I_{\mathrm{TFE}}(T_{\mathrm{max}}, F) \tau_\text{el-ph} f_{\mathrm{rep}}<I_{\mathrm{2f}}(F)$, where average current has been used for comparison with experimental $I$-$V$ data.  Given an assumed value of $\tau_\text{el-ph}$, $T_{\mathrm{max}}$ is the maximum electron temperature at the apex such that $I_{\mathrm{transient}}$ calculated in this way does not exceed the measured $I_{\mathrm{2f}}$ at any bias field.  Fig. \ref{figure:trans_bound}(a) shows $T_{\mathrm{max}}$ as a function of $\tau_\text{el-ph}$.  For $\tau_\text{el-ph} = 1$ ps, $T_{\mathrm{max}}\sim1690$~K and, for comparison, the peak apex temperature predicted by the modeling described in Section \ref{section:modeling} is well below this at 1490~K.\footnote{The 1490~K value was determined by adding the energy density deposited in the apex by a single pulse to HfC at 1150~K (determined from an $I$-$V$ fit to the data in Fig. \ref{figure:trans_bound}(a)).  The full model predicts a DC temperature of only 950~K; the experimental value may be higher due to a slight misalignment up the shank, so the 1490~K estimate is an upper limit.}  The inset to Fig. \ref{figure:trans_bound}(a) shows two examples of $I_{\mathrm{transient}}(F)$ chosen to illustrate how $T_{\mathrm{max}}$ depends on the value assumed for $\tau_\text{el-ph}$: a relatively high electron temperature is only consistent with experimental data ($I_{\mathrm{2f}}$) assuming a short $\tau_\text{el-ph}$, whereas a relatively cool temperature could emit for a much longer coupling time.  Finally, both the two-photon and transient TFE contributions to the pulsed current must be scaled to account for using parallel polarization in any experiment designed to maximize the flux from fs electron pulses.  The $\sim$30-fold increase in two-photon emission was directly measured, as in Fig. \ref{tfe2}(b), whereas modeling predicts that parallel-polarized light deposits $\sim$65\% more energy density in the apex than perpendicularly-polarized light.  Taking both scalings into account, over the useful range of bias field ($F>0.5$~GV/m where the two-photon emission turns on strongly, and $F<1.4$~GV/m, where DC TFE becomes important), we estimate that transient TFE could have contributed no more than 5\% to 25\% of the two-photon signal.  We emphasize that this is a conservative upper bound, since it corresponds to the situation shown in the inset of Fig. \ref{figure:trans_bound}(a), where the transient TFE would account for all of $I_{\mathrm{2f}}$ precisely at the first or last data point measured.

Similar ideas can be used to estimate how the transient TFE scales with laser pulse fluence in comparison with the quadratic two-photon emission scaling (Fig. \ref{figure:trans_bound}(b)).  In this experiment, the high repetition rate results in appreciable DC heating so that DC TFE dominates the current before there is appreciable transient TFE.  However, in an experiment requiring large peak current densities, the laser pulse energy could be increased while decreasing the repetition rate to avoid DC heating.  In the limit of low repetition rates, the steady-state temperature of the tip is assumed to be 300~K.  The initial temperature rise of the electron gas is estimated using an approximate energy density, $Q = \beta_Q \mathcal{F}$, with $\mathcal{F}$ the laser pulse fluence, and $\beta_Q = 220$~J/m$^3$ per nJ/cm$^2$ determined from the model of section \ref{section:modeling}.  The maximum transient electron temperature at the apex of the tip, $T_{\mathrm{trans}}$ is determined by solving
\begin{equation}
Q =\int_{T_{DC}}^{T_{trans}} C_\mathrm{el}(T)dT.
\end{equation}
As before, the normalization of the two-photon and TFE curves is determined from experimental data for our tip.  An upper limit to the average current due to transient TFE is calculated as a square pulse of duration $\tau_\text{el-ph}$, emitted from the apex with electron temperature $T_{\mathrm{trans}}$.  Explicitly, Fig. \ref{figure:trans_bound}(b) shows $I_{\mathrm{trans}} = I_{\mathrm{TFE}}(T_{\mathrm{trans}}(\mathcal{F}, F) \tau_\text{el-ph} f_{\mathrm{rep}}$ as a function of pulse fluence, for two values of bias field.  We find that transient TFE is negligible over a wide range of pulse fluences, though it may become important as the power increases to the intermediate regime ($\gamma \sim1$).  \footnote{At this point, the $P^2$ scaling of two-photon emission is expected to give way to a higher order nonlinearity, so that emission might again be dominated by prompt processes.  On the other hand, the extremely elevated transient electron temperatures that would be obtained ($\sim10^4$~K) would lead to lattice temperatures above the melting point of the tip, as well as deviations from the thermally-enhanced field emission model used here, which approximates the electronic density of states as that of a free-electron metal.}

\section{Conclusion}
We have measured two-photon emission from a hafnium carbide tip induced by sub-10~fs laser pulses.  HfC's low work function and demonstrated stability for high current density emission make it an excellent candidate tip material for ultrafast laser-triggered electron emission applications, particularly those with constraints on available peak laser intensity.  At a repetition rate of 150~MHz, the maximum current achieved is 5 electrons per pulse, limited by the onset of thermally-enhanced field emission due to an increase in the average temperature of the tip.  We estimate that for light polarized parallel to the axis of the tip under the conditions studied, transient TFE contributes negligibly to the total current.  We present a simple framework for understanding the relative scaling of multiphoton emission and thermally-enhanced field emission with changing laser parameters and bias field.

There are two complementary regimes for time-resolved pump probe measurements involving electrons.  One end of the spectrum uses large numbers of electrons in a pulse in order to achieve single-shot measurements.  At the opposite extreme, space-charge broadening may be completely eliminated by operating in a single-electron-per-pulse regime, with high repetition rates to reduce acquisition time.  Both of these regimes may be affected by the thermal effects studied here, at least in principle: at high repetition rates, the DC temperature rise of the tip can result in significant DC TFE, while at very high pulse fluence, one should consider the possibility of large temperature rises for $t<\tau_{e-\mathrm{ph}}$ and accompanying transient TFE.  The transient effect is more insidious since the resulting picosecond-timescale pulse would still be below the time resolution of standard electronic detection.  However, we show that it is not a significant limiting factor under the conditions of our high repetition rate experiment, even in HfC with its high degree of laser heating and relatively poor thermal conduction.  Comparison of tip materials shows that the favorable thermal properties of gold and tungsten greatly reduce both the steady-state temperature rise of the tip and the timescale of transient emission.  We have shown that HfC can be operated with relatively little optical power as a high repetition rate ultrafast point source of electrons in the technologically attractive regime of single electrons per pulse.

\section*{Acknowledgments}
We would like to thank Kevin Kagarice for helpful conversations and Prof. Phil Bucksbaum's group for the loan of equipment.   S. G. acknowledges support by the FWF program CoQuS (W1210-2).


\begin{thebibliography}{52}%
\makeatletter
\providecommand \@ifxundefined [1]{%
 \@ifx{#1\undefined}
}%
\providecommand \@ifnum [1]{%
 \ifnum #1\expandafter \@firstoftwo
 \else \expandafter \@secondoftwo
 \fi
}%
\providecommand \@ifx [1]{%
 \ifx #1\expandafter \@firstoftwo
 \else \expandafter \@secondoftwo
 \fi
}%
\providecommand \natexlab [1]{#1}%
\providecommand \enquote  [1]{``#1''}%
\providecommand \bibnamefont  [1]{#1}%
\providecommand \bibfnamefont [1]{#1}%
\providecommand \citenamefont [1]{#1}%
\providecommand \href@noop [0]{\@secondoftwo}%
\providecommand \href [0]{\begingroup \@sanitize@url \@href}%
\providecommand \@href[1]{\@@startlink{#1}\@@href}%
\providecommand \@@href[1]{\endgroup#1\@@endlink}%
\providecommand \@sanitize@url [0]{\catcode `\\12\catcode `\$12\catcode
  `\&12\catcode `\#12\catcode `\^12\catcode `\_12\catcode `\%12\relax}%
\providecommand \@@startlink[1]{}%
\providecommand \@@endlink[0]{}%
\providecommand \url  [0]{\begingroup\@sanitize@url \@url }%
\providecommand \@url [1]{\endgroup\@href {#1}{\urlprefix }}%
\providecommand \urlprefix  [0]{URL }%
\providecommand \Eprint [0]{\href }%
\providecommand \doibase [0]{http://dx.doi.org/}%
\providecommand \selectlanguage [0]{\@gobble}%
\providecommand \bibinfo  [0]{\@secondoftwo}%
\providecommand \bibfield  [0]{\@secondoftwo}%
\providecommand \translation [1]{[#1]}%
\providecommand \BibitemOpen [0]{}%
\providecommand \bibitemStop [0]{}%
\providecommand \bibitemNoStop [0]{.\EOS\space}%
\providecommand \EOS [0]{\spacefactor3000\relax}%
\providecommand \BibitemShut  [1]{\csname bibitem#1\endcsname}%
\let\auto@bib@innerbib\@empty
\bibitem [{\citenamefont {Spence}\ \emph {et~al.}(1994)\citenamefont {Spence},
  \citenamefont {Qian},\ and\ \citenamefont {Silverman}}]{Spence94}%
  \BibitemOpen
  \bibfield  {author} {\bibinfo {author} {\bibfnamefont {J.~C.~H.}\
  \bibnamefont {Spence}}, \bibinfo {author} {\bibfnamefont {W.}~\bibnamefont
  {Qian}}, \ and\ \bibinfo {author} {\bibfnamefont {M.~P.}\ \bibnamefont
  {Silverman}},\ }\href {\doibase 10.1116/1.579166} {\bibfield  {journal}
  {\bibinfo  {journal} {J. Vac. Sci. Technol. A}\ }\textbf {\bibinfo {volume}
  {12}},\ \bibinfo {pages} {542} (\bibinfo {year} {1994})}\BibitemShut
  {NoStop}%
\bibitem [{\citenamefont {Cho}\ \emph {et~al.}(2004)\citenamefont {Cho},
  \citenamefont {Ichimura}, \citenamefont {Shimizu},\ and\ \citenamefont
  {Oshima}}]{Cho04}%
  \BibitemOpen
  \bibfield  {author} {\bibinfo {author} {\bibfnamefont {B.}~\bibnamefont
  {Cho}}, \bibinfo {author} {\bibfnamefont {T.}~\bibnamefont {Ichimura}},
  \bibinfo {author} {\bibfnamefont {R.}~\bibnamefont {Shimizu}}, \ and\
  \bibinfo {author} {\bibfnamefont {C.}~\bibnamefont {Oshima}},\ }\href
  {\doibase 10.1103/PhysRevLett.92.246103} {\bibfield  {journal} {\bibinfo
  {journal} {Phys. Rev. Lett.}\ }\textbf {\bibinfo {volume} {92}},\ \bibinfo
  {pages} {246103} (\bibinfo {year} {2004})}\BibitemShut {NoStop}%
\bibitem [{\citenamefont {Gomer}(1961)}]{Gomer61}%
  \BibitemOpen
  \bibfield  {author} {\bibinfo {author} {\bibfnamefont {R.}~\bibnamefont
  {Gomer}},\ }\href@noop {} {\emph {\bibinfo {title} {Field Emission and Field
  Ionization}}},\ \bibinfo {series} {Harvard monographs in applied science}\
  No.~\bibinfo {number} {9}\ (\bibinfo  {publisher} {Harvard University
  Press},\ \bibinfo {address} {Cambridge},\ \bibinfo {year} {1961})\BibitemShut
  {NoStop}%
\bibitem [{\citenamefont {Midgley}\ and\ \citenamefont
  {Dunin-Borkowski}(2009)}]{Midgley09}%
  \BibitemOpen
  \bibfield  {author} {\bibinfo {author} {\bibfnamefont {P.~A.}\ \bibnamefont
  {Midgley}}\ and\ \bibinfo {author} {\bibfnamefont {R.~E.}\ \bibnamefont
  {Dunin-Borkowski}},\ }\href@noop {} {\bibfield  {journal} {\bibinfo
  {journal} {Nat. Mater.}\ }\textbf {\bibinfo {volume} {8}},\ \bibinfo {pages}
  {271} (\bibinfo {year} {2009})}\BibitemShut {NoStop}%
\bibitem [{\citenamefont {Zuo}\ \emph {et~al.}(2003)\citenamefont {Zuo},
  \citenamefont {Vartanyants}, \citenamefont {Gao}, \citenamefont {Zhang},\
  and\ \citenamefont {Nagahara}}]{Zuo03}%
  \BibitemOpen
  \bibfield  {author} {\bibinfo {author} {\bibfnamefont {J.~M.}\ \bibnamefont
  {Zuo}}, \bibinfo {author} {\bibfnamefont {I.}~\bibnamefont {Vartanyants}},
  \bibinfo {author} {\bibfnamefont {M.}~\bibnamefont {Gao}}, \bibinfo {author}
  {\bibfnamefont {R.}~\bibnamefont {Zhang}}, \ and\ \bibinfo {author}
  {\bibfnamefont {L.~A.}\ \bibnamefont {Nagahara}},\ }\href {\doibase
  10.1126/science.1083887} {\bibfield  {journal} {\bibinfo  {journal}
  {Science}\ }\textbf {\bibinfo {volume} {300}},\ \bibinfo {pages} {1419}
  (\bibinfo {year} {2003})}\BibitemShut {NoStop}%
\bibitem [{\citenamefont {Hommelhoff}\ \emph
  {et~al.}(2006{\natexlab{a}})\citenamefont {Hommelhoff}, \citenamefont
  {Sortais}, \citenamefont {Aghajani-Talesh},\ and\ \citenamefont
  {Kasevich}}]{Hommelhoff06a}%
  \BibitemOpen
  \bibfield  {author} {\bibinfo {author} {\bibfnamefont {P.}~\bibnamefont
  {Hommelhoff}}, \bibinfo {author} {\bibfnamefont {Y.}~\bibnamefont {Sortais}},
  \bibinfo {author} {\bibfnamefont {A.}~\bibnamefont {Aghajani-Talesh}}, \ and\
  \bibinfo {author} {\bibfnamefont {M.~A.}\ \bibnamefont {Kasevich}},\ }\href
  {\doibase 10.1103/PhysRevLett.96.077401} {\bibfield  {journal} {\bibinfo
  {journal} {Phys. Rev. Lett.}\ }\textbf {\bibinfo {volume} {96}},\ \bibinfo
  {pages} {077401} (\bibinfo {year} {2006}{\natexlab{a}})}\BibitemShut
  {NoStop}%
\bibitem [{\citenamefont {Hommelhoff}\ \emph
  {et~al.}(2006{\natexlab{b}})\citenamefont {Hommelhoff}, \citenamefont
  {Kealhofer},\ and\ \citenamefont {Kasevich}}]{Hommelhoff06}%
  \BibitemOpen
  \bibfield  {author} {\bibinfo {author} {\bibfnamefont {P.}~\bibnamefont
  {Hommelhoff}}, \bibinfo {author} {\bibfnamefont {C.}~\bibnamefont
  {Kealhofer}}, \ and\ \bibinfo {author} {\bibfnamefont {M.~A.}\ \bibnamefont
  {Kasevich}},\ }\href {\doibase 10.1103/PhysRevLett.97.247402} {\bibfield
  {journal} {\bibinfo  {journal} {Phys. Rev. Lett.}\ }\textbf {\bibinfo
  {volume} {97}},\ \bibinfo {pages} {247402} (\bibinfo {year}
  {2006}{\natexlab{b}})}\BibitemShut {NoStop}%
\bibitem [{\citenamefont {Ropers}\ \emph {et~al.}(2007)\citenamefont {Ropers},
  \citenamefont {Solli}, \citenamefont {Schulz}, \citenamefont {Lienau},\ and\
  \citenamefont {Elsaesser}}]{Ropers07}%
  \BibitemOpen
  \bibfield  {author} {\bibinfo {author} {\bibfnamefont {C.}~\bibnamefont
  {Ropers}}, \bibinfo {author} {\bibfnamefont {D.~R.}\ \bibnamefont {Solli}},
  \bibinfo {author} {\bibfnamefont {C.~P.}\ \bibnamefont {Schulz}}, \bibinfo
  {author} {\bibfnamefont {C.}~\bibnamefont {Lienau}}, \ and\ \bibinfo {author}
  {\bibfnamefont {T.}~\bibnamefont {Elsaesser}},\ }\href {\doibase
  10.1103/PhysRevLett.98.043907} {\bibfield  {journal} {\bibinfo  {journal}
  {Phys. Rev. Lett.}\ }\textbf {\bibinfo {volume} {98}},\ \bibinfo {pages}
  {043907} (\bibinfo {year} {2007})}\BibitemShut {NoStop}%
\bibitem [{\citenamefont {Barwick}\ \emph {et~al.}(2007)\citenamefont
  {Barwick}, \citenamefont {Corder}, \citenamefont {Strohaber}, \citenamefont
  {Chandler-Smith}, \citenamefont {Uiterwaal},\ and\ \citenamefont
  {Batelaan}}]{Barwick07}%
  \BibitemOpen
  \bibfield  {author} {\bibinfo {author} {\bibfnamefont {B.}~\bibnamefont
  {Barwick}}, \bibinfo {author} {\bibfnamefont {C.}~\bibnamefont {Corder}},
  \bibinfo {author} {\bibfnamefont {J.}~\bibnamefont {Strohaber}}, \bibinfo
  {author} {\bibfnamefont {N.}~\bibnamefont {Chandler-Smith}}, \bibinfo
  {author} {\bibfnamefont {C.}~\bibnamefont {Uiterwaal}}, \ and\ \bibinfo
  {author} {\bibfnamefont {H.}~\bibnamefont {Batelaan}},\ }\href {\doibase
  10.1088/1367-2630/9/5/142} {\bibfield  {journal} {\bibinfo  {journal} {New J.
  Phys.}\ }\textbf {\bibinfo {volume} {9}},\ \bibinfo {pages} {142} (\bibinfo
  {year} {2007})}\BibitemShut {NoStop}%
\bibitem [{\citenamefont {Ganter}\ \emph {et~al.}(2008)\citenamefont {Ganter},
  \citenamefont {Bakker}, \citenamefont {Gough}, \citenamefont {Leemann},
  \citenamefont {Paraliev}, \citenamefont {Pedrozzi}, \citenamefont {LePimpec},
  \citenamefont {Schlott}, \citenamefont {Rivkin},\ and\ \citenamefont
  {Wrulich}}]{Ganter08}%
  \BibitemOpen
  \bibfield  {author} {\bibinfo {author} {\bibfnamefont {R.}~\bibnamefont
  {Ganter}}, \bibinfo {author} {\bibfnamefont {R.}~\bibnamefont {Bakker}},
  \bibinfo {author} {\bibfnamefont {C.}~\bibnamefont {Gough}}, \bibinfo
  {author} {\bibfnamefont {S.~C.}\ \bibnamefont {Leemann}}, \bibinfo {author}
  {\bibfnamefont {M.}~\bibnamefont {Paraliev}}, \bibinfo {author}
  {\bibfnamefont {M.}~\bibnamefont {Pedrozzi}}, \bibinfo {author}
  {\bibfnamefont {F.}\ \bibnamefont {LePimpec}}, \bibinfo {author}
  {\bibfnamefont {V.}~\bibnamefont {Schlott}}, \bibinfo {author} {\bibfnamefont
  {L.}~\bibnamefont {Rivkin}}, \ and\ \bibinfo {author} {\bibfnamefont
  {A.}~\bibnamefont {Wrulich}},\ }\href {\doibase
  10.1103/PhysRevLett.100.064801} {\bibfield  {journal} {\bibinfo  {journal}
  {Phys. Rev. Lett.}\ }\textbf {\bibinfo {volume} {100}},\ \bibinfo {pages}
  {064801} (\bibinfo {year} {2008})}\BibitemShut {NoStop}%
\bibitem [{\citenamefont {Yanagisawa}\ \emph {et~al.}(2009)\citenamefont
  {Yanagisawa}, \citenamefont {Hafner}, \citenamefont {Don\'a}, \citenamefont
  {Kl\"ockner}, \citenamefont {Leuenberger}, \citenamefont {Greber},
  \citenamefont {Hengsberger},\ and\ \citenamefont
  {Osterwalder}}]{Yanagisawa10}%
  \BibitemOpen
  \bibfield  {author} {\bibinfo {author} {\bibfnamefont {H.}~\bibnamefont
  {Yanagisawa}}, \bibinfo {author} {\bibfnamefont {C.}~\bibnamefont {Hafner}},
  \bibinfo {author} {\bibfnamefont {P.}~\bibnamefont {Don\'a}}, \bibinfo
  {author} {\bibfnamefont {M.}~\bibnamefont {Kl\"ockner}}, \bibinfo {author}
  {\bibfnamefont {D.}~\bibnamefont {Leuenberger}}, \bibinfo {author}
  {\bibfnamefont {T.}~\bibnamefont {Greber}}, \bibinfo {author} {\bibfnamefont
  {M.}~\bibnamefont {Hengsberger}}, \ and\ \bibinfo {author} {\bibfnamefont
  {J.}~\bibnamefont {Osterwalder}},\ }\href {\doibase
  10.1103/PhysRevLett.103.257603} {\bibfield  {journal} {\bibinfo  {journal}
  {Phys. Rev. Lett.}\ }\textbf {\bibinfo {volume} {103}},\ \bibinfo {pages}
  {257603} (\bibinfo {year} {2009})}\BibitemShut {NoStop}%
\bibitem [{\citenamefont {Yang}\ \emph {et~al.}(2010)\citenamefont {Yang},
  \citenamefont {Mohammed},\ and\ \citenamefont {Zewail}}]{Yang10}%
  \BibitemOpen
  \bibfield  {author} {\bibinfo {author} {\bibfnamefont {D.-S.}\ \bibnamefont
  {Yang}}, \bibinfo {author} {\bibfnamefont {O.~F.}\ \bibnamefont {Mohammed}},
  \ and\ \bibinfo {author} {\bibfnamefont {A.~H.}\ \bibnamefont {Zewail}},\
  }\href {\doibase 10.1073/pnas.1009321107} {\bibfield  {journal} {\bibinfo
  {journal} {Proc. Natl. Acad. Sci. U. S. A.}\ }\textbf {\bibinfo {volume}
  {107}},\ \bibinfo {pages} {14993} (\bibinfo {year} {2010})}\BibitemShut
  {NoStop}%
\bibitem [{\citenamefont {Zewail}(2010)}]{Zewail10}%
  \BibitemOpen
  \bibfield  {author} {\bibinfo {author} {\bibfnamefont {A.~H.}\ \bibnamefont
  {Zewail}},\ }\href {\doibase 10.1126/science.1166135} {\bibfield  {journal}
  {\bibinfo  {journal} {Science}\ }\textbf {\bibinfo {volume} {328}},\ \bibinfo
  {pages} {187} (\bibinfo {year} {2010})}\BibitemShut {NoStop}%
\bibitem [{\citenamefont {Miller}\ \emph {et~al.}(2010)\citenamefont {Miller},
  \citenamefont {Ernstorfer}, \citenamefont {Harb}, \citenamefont {Gao},
  \citenamefont {Hebeisen}, \citenamefont {Jean-Ruel}, \citenamefont {Lu},
  \citenamefont {Moriena},\ and\ \citenamefont {Sciaini}}]{Miller10}%
  \BibitemOpen
  \bibfield  {author} {\bibinfo {author} {\bibfnamefont {R.}~\bibnamefont
  {Miller}}, \bibinfo {author} {\bibfnamefont {R.}~\bibnamefont {Ernstorfer}},
  \bibinfo {author} {\bibfnamefont {M.}~\bibnamefont {Harb}}, \bibinfo {author}
  {\bibfnamefont {M.}~\bibnamefont {Gao}}, \bibinfo {author} {\bibfnamefont
  {C.}~\bibnamefont {Hebeisen}}, \bibinfo {author} {\bibfnamefont
  {H.}~\bibnamefont {Jean-Ruel}}, \bibinfo {author} {\bibfnamefont
  {C.}~\bibnamefont {Lu}}, \bibinfo {author} {\bibfnamefont {G.}~\bibnamefont
  {Moriena}}, \ and\ \bibinfo {author} {\bibfnamefont {G.}~\bibnamefont
  {Sciaini}},\ }\href {\doibase 10.1107/S0108767309053926} {\bibfield
  {journal} {\bibinfo  {journal} {Acta Crystallogr., Sect. A}\ }\textbf
  {\bibinfo {volume} {66}},\ \bibinfo {pages} {137} (\bibinfo {year}
  {2010})}\BibitemShut {NoStop}%
\bibitem [{\citenamefont {Hasselbach}(2010)}]{Hasselbach10}%
  \BibitemOpen
  \bibfield  {author} {\bibinfo {author} {\bibfnamefont {F.}~\bibnamefont
  {Hasselbach}},\ }\href {http://stacks.iop.org/0034-4885/73/i=1/a=016101}
  {\bibfield  {journal} {\bibinfo  {journal} {Reports on Progress in Physics}\
  }\textbf {\bibinfo {volume} {73}},\ \bibinfo {pages} {016101} (\bibinfo
  {year} {2010})}\BibitemShut {NoStop}%
\bibitem [{\citenamefont {Bormann}\ \emph {et~al.}(2010)\citenamefont
  {Bormann}, \citenamefont {Gulde}, \citenamefont {Weismann}, \citenamefont
  {Yalunin},\ and\ \citenamefont {Ropers}}]{Bormann10}%
  \BibitemOpen
  \bibfield  {author} {\bibinfo {author} {\bibfnamefont {R.}~\bibnamefont
  {Bormann}}, \bibinfo {author} {\bibfnamefont {M.}~\bibnamefont {Gulde}},
  \bibinfo {author} {\bibfnamefont {A.}~\bibnamefont {Weismann}}, \bibinfo
  {author} {\bibfnamefont {S.~V.}\ \bibnamefont {Yalunin}}, \ and\ \bibinfo
  {author} {\bibfnamefont {C.}~\bibnamefont {Ropers}},\ }\href {\doibase
  10.1103/PhysRevLett.105.147601} {\bibfield  {journal} {\bibinfo  {journal}
  {Phys. Rev. Lett.}\ }\textbf {\bibinfo {volume} {105}},\ \bibinfo {pages}
  {147601} (\bibinfo {year} {2010})}\BibitemShut {NoStop}%
\bibitem [{\citenamefont {Schenk}\ \emph {et~al.}(2010)\citenamefont {Schenk},
  \citenamefont {Kr\"uger},\ and\ \citenamefont {Hommelhoff}}]{Schenk10}%
  \BibitemOpen
  \bibfield  {author} {\bibinfo {author} {\bibfnamefont {M.}~\bibnamefont
  {Schenk}}, \bibinfo {author} {\bibfnamefont {M.}~\bibnamefont {Kr\"uger}}, \
  and\ \bibinfo {author} {\bibfnamefont {P.}~\bibnamefont {Hommelhoff}},\
  }\href {\doibase 10.1103/PhysRevLett.105.257601} {\bibfield  {journal}
  {\bibinfo  {journal} {Phys. Rev. Lett.}\ }\textbf {\bibinfo {volume} {105}},\
  \bibinfo {pages} {257601} (\bibinfo {year} {2010})}\BibitemShut {NoStop}%
\bibitem [{\citenamefont {Caprez}\ \emph {et~al.}(2007)\citenamefont {Caprez},
  \citenamefont {Barwick},\ and\ \citenamefont {Batelaan}}]{Caprez07}%
  \BibitemOpen
  \bibfield  {author} {\bibinfo {author} {\bibfnamefont {A.}~\bibnamefont
  {Caprez}}, \bibinfo {author} {\bibfnamefont {B.}~\bibnamefont {Barwick}}, \
  and\ \bibinfo {author} {\bibfnamefont {H.}~\bibnamefont {Batelaan}},\ }\href
  {\doibase 10.1103/PhysRevLett.99.210401} {\bibfield  {journal} {\bibinfo
  {journal} {Phys. Rev. Lett.}\ }\textbf {\bibinfo {volume} {99}},\ \bibinfo
  {pages} {210401} (\bibinfo {year} {2007})}\BibitemShut {NoStop}%
\bibitem [{\citenamefont {Plettner}\ \emph {et~al.}(2006)\citenamefont
  {Plettner}, \citenamefont {Lu},\ and\ \citenamefont {Byer}}]{Plettner06}%
  \BibitemOpen
  \bibfield  {author} {\bibinfo {author} {\bibfnamefont {T.}~\bibnamefont
  {Plettner}}, \bibinfo {author} {\bibfnamefont {P.~P.}\ \bibnamefont {Lu}}, \
  and\ \bibinfo {author} {\bibfnamefont {R.~L.}\ \bibnamefont {Byer}},\
  }\href@noop {} {\bibfield  {journal} {\bibinfo  {journal} {Phys. Rev. Spec.
  Top. Accel. Beams}\ }\textbf {\bibinfo {volume} {9}},\ \bibinfo {pages}
  {111301} (\bibinfo {year} {2006})}\BibitemShut {NoStop}%
\bibitem [{\citenamefont {Kagarice}\ \emph {et~al.}(2008)\citenamefont
  {Kagarice}, \citenamefont {Magera}, \citenamefont {Pollard},\ and\
  \citenamefont {Mackie}}]{Kagarice08}%
  \BibitemOpen
  \bibfield  {author} {\bibinfo {author} {\bibfnamefont {K.~J.}\ \bibnamefont
  {Kagarice}}, \bibinfo {author} {\bibfnamefont {G.~G.}\ \bibnamefont
  {Magera}}, \bibinfo {author} {\bibfnamefont {S.~D.}\ \bibnamefont {Pollard}},
  \ and\ \bibinfo {author} {\bibfnamefont {W.~A.}\ \bibnamefont {Mackie}},\
  }\href {\doibase 10.1116/1.2812535} {\bibfield  {journal} {\bibinfo
  {journal} {J. Vac. Sci. Technol. B}\ }\textbf {\bibinfo {volume} {26}},\
  \bibinfo {pages} {868} (\bibinfo {year} {2008})}\BibitemShut {NoStop}%
\bibitem [{\citenamefont {Mackie}\ \emph {et~al.}(1993)\citenamefont {Mackie},
  \citenamefont {Hartman},\ and\ \citenamefont {Davis}}]{Mackie93}%
  \BibitemOpen
  \bibfield  {author} {\bibinfo {author} {\bibfnamefont {W.~A.}\ \bibnamefont
  {Mackie}}, \bibinfo {author} {\bibfnamefont {R.~L.}\ \bibnamefont {Hartman}},
  \ and\ \bibinfo {author} {\bibfnamefont {P.~R.}\ \bibnamefont {Davis}},\
  }\href {\doibase 10.1016/0169-4332(93)90290-R} {\bibfield  {journal}
  {\bibinfo  {journal} {Appl. Surf. Sci.}\ }\textbf {\bibinfo {volume} {67}},\
  \bibinfo {pages} {29} (\bibinfo {year} {1993})}\BibitemShut {NoStop}%
\bibitem [{\citenamefont {Vurpillot}\ \emph {et~al.}(2009)\citenamefont
  {Vurpillot}, \citenamefont {Houard}, \citenamefont {Vella},\ and\
  \citenamefont {Deconihout}}]{Vurpillot09}%
  \BibitemOpen
  \bibfield  {author} {\bibinfo {author} {\bibfnamefont {F.}~\bibnamefont
  {Vurpillot}}, \bibinfo {author} {\bibfnamefont {J.}~\bibnamefont {Houard}},
  \bibinfo {author} {\bibfnamefont {A.}~\bibnamefont {Vella}}, \ and\ \bibinfo
  {author} {\bibfnamefont {B.}~\bibnamefont {Deconihout}},\ }\href
  {http://stacks.iop.org/0022-3727/42/i=12/a=125502} {\bibfield  {journal}
  {\bibinfo  {journal} {J. Phys. D: Appl. Phys.}\ }\textbf {\bibinfo {volume}
  {42}},\ \bibinfo {pages} {125502} (\bibinfo {year} {2009})}\BibitemShut
  {NoStop}%
\bibitem [{\citenamefont {Houard}\ \emph {et~al.}(2011)\citenamefont {Houard},
  \citenamefont {Vella}, \citenamefont {Vurpillot},\ and\ \citenamefont
  {Deconihout}}]{Houard11}%
  \BibitemOpen
  \bibfield  {author} {\bibinfo {author} {\bibfnamefont {J.}~\bibnamefont
  {Houard}}, \bibinfo {author} {\bibfnamefont {A.}~\bibnamefont {Vella}},
  \bibinfo {author} {\bibfnamefont {F.}~\bibnamefont {Vurpillot}}, \ and\
  \bibinfo {author} {\bibfnamefont {B.}~\bibnamefont {Deconihout}},\ }\href
  {\doibase 10.1103/PhysRevB.84.033405} {\bibfield  {journal} {\bibinfo
  {journal} {Phys. Rev. B}\ }\textbf {\bibinfo {volume} {84}},\ \bibinfo
  {pages} {033405} (\bibinfo {year} {2011})}\BibitemShut {NoStop}%
\bibitem [{\citenamefont {Zaima}\ \emph {et~al.}(1984)\citenamefont {Zaima},
  \citenamefont {Adachi},\ and\ \citenamefont {Shibata}}]{Zaima84}%
  \BibitemOpen
  \bibfield  {author} {\bibinfo {author} {\bibfnamefont {S.}~\bibnamefont
  {Zaima}}, \bibinfo {author} {\bibfnamefont {H.}~\bibnamefont {Adachi}}, \
  and\ \bibinfo {author} {\bibfnamefont {Y.}~\bibnamefont {Shibata}},\ }\href
  {\doibase 10.1116/1.582919} {\bibfield  {journal} {\bibinfo  {journal} {J.
  Vac. Sci. Technol. B}\ }\textbf {\bibinfo {volume} {2}},\ \bibinfo {pages}
  {73} (\bibinfo {year} {1984})}\BibitemShut {NoStop}%
\bibitem [{\citenamefont {Murphy}\ and\ \citenamefont {Good}(1956)}]{Murphy56}%
  \BibitemOpen
  \bibfield  {author} {\bibinfo {author} {\bibfnamefont {E.~L.}\ \bibnamefont
  {Murphy}}\ and\ \bibinfo {author} {\bibfnamefont {R.~H.}\ \bibnamefont
  {Good}},\ }\href@noop {} {\bibfield  {journal} {\bibinfo  {journal} {Physical
  Review}\ }\textbf {\bibinfo {volume} {102}},\ \bibinfo {pages} {1464}
  (\bibinfo {year} {1956})}\BibitemShut {NoStop}%
\bibitem [{\citenamefont {Girardeau-Montaut}\ and\ \citenamefont
  {Girardeau-Montaut}(1995)}]{Girardeau-Montaut95}%
  \BibitemOpen
  \bibfield  {author} {\bibinfo {author} {\bibfnamefont {J.~P.}\ \bibnamefont
  {Girardeau-Montaut}}\ and\ \bibinfo {author} {\bibfnamefont {C.}~\bibnamefont
  {Girardeau-Montaut}},\ }\href {\doibase 10.1103/PhysRevB.51.13560} {\bibfield
   {journal} {\bibinfo  {journal} {Phys. Rev. B}\ }\textbf {\bibinfo {volume}
  {51}},\ \bibinfo {pages} {13560} (\bibinfo {year} {1995})}\BibitemShut
  {NoStop}%
\bibitem [{\citenamefont {Lee}(1973)}]{Lee73}%
  \BibitemOpen
  \bibfield  {author} {\bibinfo {author} {\bibfnamefont {M.~J.~G.}\
  \bibnamefont {Lee}},\ }\href@noop {} {\bibfield  {journal} {\bibinfo
  {journal} {Phys. Rev. Lett.}\ }\textbf {\bibinfo {volume} {30}},\ \bibinfo
  {pages} {1193} (\bibinfo {year} {1973})}\BibitemShut {NoStop}%
\bibitem [{\citenamefont {Mainfray}\ and\ \citenamefont
  {Manus}(1991)}]{Mainfray91}%
  \BibitemOpen
  \bibfield  {author} {\bibinfo {author} {\bibfnamefont {G.}~\bibnamefont
  {Mainfray}}\ and\ \bibinfo {author} {\bibfnamefont {C.}~\bibnamefont
  {Manus}},\ }\href {\doibase 10.1088/0034-4885/54/10/002} {\bibfield
  {journal} {\bibinfo  {journal} {Rep. Prog. Phys.}\ }\textbf {\bibinfo
  {volume} {54}},\ \bibinfo {pages} {1333} (\bibinfo {year}
  {1991})}\BibitemShut {NoStop}%
\bibitem [{\citenamefont {Brabec}\ and\ \citenamefont
  {Krausz}(2000)}]{Brabec00}%
  \BibitemOpen
  \bibfield  {author} {\bibinfo {author} {\bibfnamefont {T.}~\bibnamefont
  {Brabec}}\ and\ \bibinfo {author} {\bibfnamefont {F.}~\bibnamefont
  {Krausz}},\ }\href {\doibase 10.1103/RevModPhys.72.545} {\bibfield  {journal}
  {\bibinfo  {journal} {Rev. Mod. Phys.}\ }\textbf {\bibinfo {volume} {72}},\
  \bibinfo {pages} {545} (\bibinfo {year} {2000})}\BibitemShut {NoStop}%
\bibitem [{\citenamefont {Yudin}\ and\ \citenamefont {Ivanov}(2001)}]{Yudin01}%
  \BibitemOpen
  \bibfield  {author} {\bibinfo {author} {\bibfnamefont {G.~L.}\ \bibnamefont
  {Yudin}}\ and\ \bibinfo {author} {\bibfnamefont {M.~Y.}\ \bibnamefont
  {Ivanov}},\ }\href {\doibase 10.1103/PhysRevA.64.013409} {\bibfield
  {journal} {\bibinfo  {journal} {Phys. Rev. A}\ }\textbf {\bibinfo {volume}
  {64}},\ \bibinfo {pages} {013409} (\bibinfo {year} {2001})}\BibitemShut
  {NoStop}%
\bibitem [{\citenamefont {Uiberacker}\ \emph {et~al.}(2007)\citenamefont
  {Uiberacker}, \citenamefont {Uphues}, \citenamefont {Schultze}, \citenamefont
  {Verhoef}, \citenamefont {Yakovlev}, \citenamefont {Kling}, \citenamefont
  {Rauschenberger}, \citenamefont {Kabachnik}, \citenamefont {Schr\"{o}der},
  \citenamefont {Lezius}, \citenamefont {Kompa}, \citenamefont {Muller},
  \citenamefont {Vrakking}, \citenamefont {Hendel}, \citenamefont {Kleineberg},
  \citenamefont {Heinzmann}, \citenamefont {Drescher},\ and\ \citenamefont
  {Krausz}}]{Uiberacker07}%
  \BibitemOpen
  \bibfield  {author} {\bibinfo {author} {\bibfnamefont {M.}~\bibnamefont
  {Uiberacker}}, \bibinfo {author} {\bibfnamefont {T.}~\bibnamefont {Uphues}},
  \bibinfo {author} {\bibfnamefont {M.}~\bibnamefont {Schultze}}, \bibinfo
  {author} {\bibfnamefont {A.~J.}\ \bibnamefont {Verhoef}}, \bibinfo {author}
  {\bibfnamefont {V.}~\bibnamefont {Yakovlev}}, \bibinfo {author}
  {\bibfnamefont {M.~F.}\ \bibnamefont {Kling}}, \bibinfo {author}
  {\bibfnamefont {J.}~\bibnamefont {Rauschenberger}}, \bibinfo {author}
  {\bibfnamefont {N.~M.}\ \bibnamefont {Kabachnik}}, \bibinfo {author}
  {\bibfnamefont {H.}~\bibnamefont {Schr\"{o}der}}, \bibinfo {author}
  {\bibfnamefont {M.}~\bibnamefont {Lezius}}, \bibinfo {author} {\bibfnamefont
  {K.~L.}\ \bibnamefont {Kompa}}, \bibinfo {author} {\bibfnamefont {H.-G.}\
  \bibnamefont {Muller}}, \bibinfo {author} {\bibfnamefont {M.~J.~J.}\
  \bibnamefont {Vrakking}}, \bibinfo {author} {\bibfnamefont {S.}~\bibnamefont
  {Hendel}}, \bibinfo {author} {\bibfnamefont {U.}~\bibnamefont {Kleineberg}},
  \bibinfo {author} {\bibfnamefont {U.}~\bibnamefont {Heinzmann}}, \bibinfo
  {author} {\bibfnamefont {M.}~\bibnamefont {Drescher}}, \ and\ \bibinfo
  {author} {\bibfnamefont {F.}~\bibnamefont {Krausz}},\ }\href {\doibase
  10.1038/nature05648} {\bibfield  {journal} {\bibinfo  {journal} {Nature}\
  }\textbf {\bibinfo {volume} {446}},\ \bibinfo {pages} {627} (\bibinfo {year}
  {2007})}\BibitemShut {NoStop}%
\bibitem [{\citenamefont {Anisimov}\ \emph {et~al.}(1974)\citenamefont
  {Anisimov}, \citenamefont {Kapeliovich},\ and\ \citenamefont
  {Perl'man}}]{Anisimov74}%
  \BibitemOpen
  \bibfield  {author} {\bibinfo {author} {\bibfnamefont {S.~I.}\ \bibnamefont
  {Anisimov}}, \bibinfo {author} {\bibfnamefont {B.~L.}\ \bibnamefont
  {Kapeliovich}}, \ and\ \bibinfo {author} {\bibfnamefont {T.~L.}\ \bibnamefont
  {Perl'man}},\ }\href@noop {} {\bibfield  {journal} {\bibinfo  {journal} {Sov.
  Phys. - JETP}\ }\textbf {\bibinfo {volume} {39}},\ \bibinfo {pages} {375}
  (\bibinfo {year} {1974})}\BibitemShut {NoStop}%
\bibitem [{\citenamefont {Lee}\ and\ \citenamefont {Robins}(1989)}]{Lee89}%
  \BibitemOpen
  \bibfield  {author} {\bibinfo {author} {\bibfnamefont {M.~J.~G.}\
  \bibnamefont {Lee}}\ and\ \bibinfo {author} {\bibfnamefont {E.~S.}\
  \bibnamefont {Robins}},\ }\href {\doibase 10.1063/1.342941} {\bibfield
  {journal} {\bibinfo  {journal} {J. Appl. Phys.}\ }\textbf {\bibinfo {volume}
  {65}},\ \bibinfo {pages} {1699} (\bibinfo {year} {1989})}\BibitemShut
  {NoStop}%
\bibitem [{\citenamefont {Venus}\ and\ \citenamefont {Lee}(1983)}]{Venus83}%
  \BibitemOpen
  \bibfield  {author} {\bibinfo {author} {\bibfnamefont {D.}~\bibnamefont
  {Venus}}\ and\ \bibinfo {author} {\bibfnamefont {M.}~\bibnamefont {Lee}},\
  }\href {\doibase DOI: 10.1016/0039-6028(83)90577-0} {\bibfield  {journal}
  {\bibinfo  {journal} {Surf. Sci.}\ }\textbf {\bibinfo {volume} {125}},\
  \bibinfo {pages} {452 } (\bibinfo {year} {1983})}\BibitemShut {NoStop}%
\bibitem [{\citenamefont {Lee}\ \emph {et~al.}(1980)\citenamefont {Lee},
  \citenamefont {Reifenberger}, \citenamefont {Robins},\ and\ \citenamefont
  {Lindenmayr}}]{Lee80}%
  \BibitemOpen
  \bibfield  {author} {\bibinfo {author} {\bibfnamefont {M.~J.~G.}\
  \bibnamefont {Lee}}, \bibinfo {author} {\bibfnamefont {R.}~\bibnamefont
  {Reifenberger}}, \bibinfo {author} {\bibfnamefont {E.~S.}\ \bibnamefont
  {Robins}}, \ and\ \bibinfo {author} {\bibfnamefont {H.~G.}\ \bibnamefont
  {Lindenmayr}},\ }\href {\doibase 10.1063/1.328379} {\bibfield  {journal}
  {\bibinfo  {journal} {J. Appl. Phys.}\ }\textbf {\bibinfo {volume} {51}},\
  \bibinfo {pages} {4996} (\bibinfo {year} {1980})}\BibitemShut {NoStop}%
\bibitem [{\citenamefont {using the software package Comsol~version
  3.5a}(2009)}]{comsol3.5a}%
  \BibitemOpen
  \bibfield  {author} {\bibinfo {author} {\bibnamefont {using the software
  package Comsol~version 3.5a}},\ }\href@noop {} {} (\bibinfo {year} {Comsol
  AB, Sweden, 2009})\BibitemShut {NoStop}%
\bibitem [{\citenamefont {Pierson}(1996)}]{Pierson96}%
  \BibitemOpen
  \bibfield  {author} {\bibinfo {author} {\bibfnamefont {H.~O.}\ \bibnamefont
  {Pierson}},\ }\href@noop {} {\emph {\bibinfo {title} {Handbook of Refractory
  Carbides and Nitrides}}}\ (\bibinfo  {publisher} {Westwood, N. J.: Noyes
  Publications},\ \bibinfo {year} {1996})\BibitemShut {NoStop}%
\bibitem [{\citenamefont {Linstrom}\ and\ \citenamefont
  {Mallard}()}]{NISTChemWebBook}%
  \BibitemOpen
  \bibinfo {editor} {\bibfnamefont {P.~J.}\ \bibnamefont {Linstrom}}\ and\
  \bibinfo {editor} {\bibfnamefont {W.~G.}\ \bibnamefont {Mallard}},\ eds.,\
  \href {http://webbook.nist.gov} {\emph {\bibinfo {title} {NIST Chemistry
  WebBook, NIST Standard Reference Database Number 69}}}\ (\bibinfo
  {publisher} {National Institute of Standards and Technology},\ \bibinfo
  {address} {Gaithersburg MD, 20899})\BibitemShut {NoStop}%
\bibitem [{\citenamefont {Takahashi}\ and\ \citenamefont
  {Akiyama}(1986)}]{Takahashi86}%
  \BibitemOpen
  \bibfield  {author} {\bibinfo {author} {\bibfnamefont {Y.}~\bibnamefont
  {Takahashi}}\ and\ \bibinfo {author} {\bibfnamefont {H.}~\bibnamefont
  {Akiyama}},\ }\href {\doibase 10.1016/0040-6031(86)85012-2} {\bibfield
  {journal} {\bibinfo  {journal} {Thermochimica Acta}\ }\textbf {\bibinfo
  {volume} {109}},\ \bibinfo {pages} {105 } (\bibinfo {year}
  {1986})}\BibitemShut {NoStop}%
\bibitem [{\citenamefont {Ihara}\ \emph {et~al.}(1976)\citenamefont {Ihara},
  \citenamefont {Hirabayashi},\ and\ \citenamefont {Nakagawa}}]{Ihara76}%
  \BibitemOpen
  \bibfield  {author} {\bibinfo {author} {\bibfnamefont {H.}~\bibnamefont
  {Ihara}}, \bibinfo {author} {\bibfnamefont {M.}~\bibnamefont {Hirabayashi}},
  \ and\ \bibinfo {author} {\bibfnamefont {H.}~\bibnamefont {Nakagawa}},\
  }\href {\doibase 10.1103/PhysRevB.14.1707} {\bibfield  {journal} {\bibinfo
  {journal} {Phys. Rev. B}\ }\textbf {\bibinfo {volume} {14}},\ \bibinfo
  {pages} {1707} (\bibinfo {year} {1976})}\BibitemShut {NoStop}%
\bibitem [{\citenamefont {Lin}\ \emph {et~al.}(2008)\citenamefont {Lin},
  \citenamefont {Zhigilei},\ and\ \citenamefont {Celli}}]{Lin08}%
  \BibitemOpen
  \bibfield  {author} {\bibinfo {author} {\bibfnamefont {Z.}~\bibnamefont
  {Lin}}, \bibinfo {author} {\bibfnamefont {L.~V.}\ \bibnamefont {Zhigilei}}, \
  and\ \bibinfo {author} {\bibfnamefont {V.}~\bibnamefont {Celli}},\ }\href
  {\doibase 10.1103/PhysRevB.77.075133} {\bibfield  {journal} {\bibinfo
  {journal} {Phys. Rev. B}\ }\textbf {\bibinfo {volume} {77}},\ \bibinfo
  {pages} {075133} (\bibinfo {year} {2008})}\BibitemShut {NoStop}%
\bibitem [{\citenamefont {Wuchina}\ \emph {et~al.}(2004)\citenamefont
  {Wuchina}, \citenamefont {Opeka}, \citenamefont {Causey}, \citenamefont
  {Buesking}, \citenamefont {Spain}, \citenamefont {Cull}, \citenamefont
  {Routbort},\ and\ \citenamefont {Guitierrez-Mora}}]{Wuchina04}%
  \BibitemOpen
  \bibfield  {author} {\bibinfo {author} {\bibfnamefont {E.}~\bibnamefont
  {Wuchina}}, \bibinfo {author} {\bibfnamefont {M.}~\bibnamefont {Opeka}},
  \bibinfo {author} {\bibfnamefont {S.}~\bibnamefont {Causey}}, \bibinfo
  {author} {\bibfnamefont {K.}~\bibnamefont {Buesking}}, \bibinfo {author}
  {\bibfnamefont {J.}~\bibnamefont {Spain}}, \bibinfo {author} {\bibfnamefont
  {A.}~\bibnamefont {Cull}}, \bibinfo {author} {\bibfnamefont {J.}~\bibnamefont
  {Routbort}}, \ and\ \bibinfo {author} {\bibfnamefont {F.}~\bibnamefont
  {Guitierrez-Mora}},\ }\href
  {http://dx.doi.org/10.1023/B:JMSC.0000041690.06117.34} {\bibfield  {journal}
  {\bibinfo  {journal} {J. Mater. Sci.}\ }\textbf {\bibinfo {volume} {39}},\
  \bibinfo {pages} {5939} (\bibinfo {year} {2004})}\BibitemShut {NoStop}%
\bibitem [{\citenamefont {Haynes}\ and\ \citenamefont {Lide}(2011)}]{CRC92}%
  \BibitemOpen
  \bibinfo {editor} {\bibfnamefont {W.~M.}\ \bibnamefont {Haynes}}\ and\
  \bibinfo {editor} {\bibfnamefont {D.~R.}\ \bibnamefont {Lide}},\ eds.,\
  \href@noop {} {\emph {\bibinfo {title} {CRC Handbook of Chemistry and
  Physics}}},\ \bibinfo {edition} {92nd}\ ed.\ (\bibinfo  {publisher} {CRC
  Press},\ \bibinfo {address} {Boca Raton, FL},\ \bibinfo {year}
  {2011})\BibitemShut {NoStop}%
\bibitem [{\citenamefont {Christensen}\ \emph {et~al.}(2007)\citenamefont
  {Christensen}, \citenamefont {Vestentoft},\ and\ \citenamefont
  {Balling}}]{Christensen07}%
  \BibitemOpen
  \bibfield  {author} {\bibinfo {author} {\bibfnamefont {B.}~\bibnamefont
  {Christensen}}, \bibinfo {author} {\bibfnamefont {K.}~\bibnamefont
  {Vestentoft}}, \ and\ \bibinfo {author} {\bibfnamefont {P.}~\bibnamefont
  {Balling}},\ }\href {\doibase 10.1016/j.apsusc.2007.01.045} {\bibfield
  {journal} {\bibinfo  {journal} {Applied Surface Science}\ }\textbf {\bibinfo
  {volume} {253}},\ \bibinfo {pages} {6347 } (\bibinfo {year} {2007})},\
  \bibinfo {note} {proceedings of the Fifth International Conference on
  Photo-Excited Processes and Applications}\BibitemShut {NoStop}%
\bibitem [{\citenamefont {Fl\"{u}ck}\ \emph {et~al.}(1982)\citenamefont
  {Fl\"{u}ck}, \citenamefont {Geserich},\ and\ \citenamefont
  {Politis}}]{Fluck82}%
  \BibitemOpen
  \bibfield  {author} {\bibinfo {author} {\bibfnamefont {F.~W.}\ \bibnamefont
  {Fl\"{u}ck}}, \bibinfo {author} {\bibfnamefont {H.~P.}\ \bibnamefont
  {Geserich}}, \ and\ \bibinfo {author} {\bibfnamefont {C.}~\bibnamefont
  {Politis}},\ }\href {\doibase 10.1002/pssb.2221120122} {\bibfield  {journal}
  {\bibinfo  {journal} {Phys. Status Solidi B}\ }\textbf {\bibinfo {volume}
  {112}},\ \bibinfo {pages} {193} (\bibinfo {year} {1982})}\BibitemShut
  {NoStop}%
\bibitem [{\citenamefont {Johnson}\ and\ \citenamefont
  {Christy}(1972)}]{Johnson72}%
  \BibitemOpen
  \bibfield  {author} {\bibinfo {author} {\bibfnamefont {P.~B.}\ \bibnamefont
  {Johnson}}\ and\ \bibinfo {author} {\bibfnamefont {R.~W.}\ \bibnamefont
  {Christy}},\ }\href {\doibase 10.1103/PhysRevB.6.4370} {\bibfield  {journal}
  {\bibinfo  {journal} {Phys. Rev. B}\ }\textbf {\bibinfo {volume} {6}},\
  \bibinfo {pages} {4370} (\bibinfo {year} {1972})}\BibitemShut {NoStop}%
\bibitem [{\citenamefont {Martin}\ \emph {et~al.}(2001)\citenamefont {Martin},
  \citenamefont {Hamann},\ and\ \citenamefont {Wickramasinghe}}]{Martin01}%
  \BibitemOpen
  \bibfield  {author} {\bibinfo {author} {\bibfnamefont {Y.~C.}\ \bibnamefont
  {Martin}}, \bibinfo {author} {\bibfnamefont {H.~F.}\ \bibnamefont {Hamann}},
  \ and\ \bibinfo {author} {\bibfnamefont {H.~K.}\ \bibnamefont
  {Wickramasinghe}},\ }\href {\doibase 10.1063/1.1354655} {\bibfield  {journal}
  {\bibinfo  {journal} {J. Appl. Phys.}\ }\textbf {\bibinfo {volume} {89}},\
  \bibinfo {pages} {5774} (\bibinfo {year} {2001})}\BibitemShut {NoStop}%
\bibitem [{\citenamefont {Yanagisawa}\ \emph {et~al.}(2010)\citenamefont
  {Yanagisawa}, \citenamefont {Hafner}, \citenamefont {Don\'a}, \citenamefont
  {Kl\"ockner}, \citenamefont {Leuenberger}, \citenamefont {Greber},
  \citenamefont {Osterwalder},\ and\ \citenamefont
  {Hengsberger}}]{Yanagisawa10a}%
  \BibitemOpen
  \bibfield  {author} {\bibinfo {author} {\bibfnamefont {H.}~\bibnamefont
  {Yanagisawa}}, \bibinfo {author} {\bibfnamefont {C.}~\bibnamefont {Hafner}},
  \bibinfo {author} {\bibfnamefont {P.}~\bibnamefont {Don\'a}}, \bibinfo
  {author} {\bibfnamefont {M.}~\bibnamefont {Kl\"ockner}}, \bibinfo {author}
  {\bibfnamefont {D.}~\bibnamefont {Leuenberger}}, \bibinfo {author}
  {\bibfnamefont {T.}~\bibnamefont {Greber}}, \bibinfo {author} {\bibfnamefont
  {J.}~\bibnamefont {Osterwalder}}, \ and\ \bibinfo {author} {\bibfnamefont
  {M.}~\bibnamefont {Hengsberger}},\ }\href {\doibase
  10.1103/PhysRevB.81.115429} {\bibfield  {journal} {\bibinfo  {journal} {Phys.
  Rev. B}\ }\textbf {\bibinfo {volume} {81}},\ \bibinfo {pages} {115429}
  (\bibinfo {year} {2010})}\BibitemShut {NoStop}%
\bibitem [{\citenamefont {Vestentoft}\ and\ \citenamefont
  {Balling}(2006)}]{Vestentoft06}%
  \BibitemOpen
  \bibfield  {author} {\bibinfo {author} {\bibfnamefont {K.}~\bibnamefont
  {Vestentoft}}\ and\ \bibinfo {author} {\bibfnamefont {P.}~\bibnamefont
  {Balling}},\ }\href {http://dx.doi.org/10.1007/s00339-006-3602-4} {\bibfield
  {journal} {\bibinfo  {journal} {Applied Physics A: Materials Science \&
  Processing}\ }\textbf {\bibinfo {volume} {84}},\ \bibinfo {pages} {207}
  (\bibinfo {year} {2006})},\ \bibinfo {note}
  {10.1007/s00339-006-3602-4}\BibitemShut {NoStop}%
\bibitem [{\citenamefont {Wellershoff}(1999)}]{Wellershoff99}%
  \BibitemOpen
  \bibfield  {author} {\bibinfo {author} {\bibfnamefont {J.~H. J. G. E.~M.}\
  \bibnamefont {Wellershoff}, \bibfnamefont {S.-S.}},\ }\href
  {http://dx.doi.org/10.1007/s003399900305} {\bibfield  {journal} {\bibinfo
  {journal} {Applied Physics A: Materials Science \&amp; Processing}\ }\textbf
  {\bibinfo {volume} {69}},\ \bibinfo {pages} {S99} (\bibinfo {year}
  {1999})}\BibitemShut {NoStop}%
\bibitem [{Note1()}]{Note1}%
  \BibitemOpen
  \bibinfo {note} {The 1490~K value was determined by adding the energy density
  deposited in the apex by a single pulse to HfC at 1150~K (determined from an
  $I$-$V$ fit to the data in Fig. \ref {figure:trans_bound}(a)). The full model
  predicts a DC temperature of only 950~K; the experimental value may be higher
  due to a slight misalignment up the shank, so the 1490~K estimate is an upper
  limit.}\BibitemShut {Stop}%
\bibitem [{Note2()}]{Note2}%
  \BibitemOpen
  \bibinfo {note} {At this point, the $P^2$ scaling of two-photon emission is
  expected to give way to a higher order nonlinearity, so that emission might
  again be dominated by prompt processes. On the other hand, the extremely
  elevated transient electron temperatures that would be obtained ($\sim
  10^4$~K) would lead to lattice temperatures above the melting point of the
  tip, as well as deviations from the thermally-enhanced field emission model
  used here, which approximates the electronic density of states as that of a
  free-electron metal.}\BibitemShut {Stop}%
\end{thebibliography}
\end{document}